\documentclass[fleqn,usenatbib,useAMS]{mnras}

\usepackage{graphicx}	
\graphicspath{{Figures/}} 
\usepackage{amsmath}	
\usepackage{amssymb}	
\usepackage{multicol}        
\usepackage{bm}		
\usepackage{pdflscape}	
\usepackage{physics}
\newcommand{\dunderline}[1]{\underline{\underline{#1}}}

\usepackage[T1]{fontenc}
\usepackage{ae,aecompl}

\usepackage{newtxtext,newtxmath}

\usepackage{color}
\definecolor{red}{RGB}{0, 0, 0}
\newcommand{\edit}[1]{\textcolor{red}{#1}}

\title[A resistive extension for ideal MHD]{A resistive extension for ideal MHD}

\author[A. J. Wright \& I. Hawke]{Alex James Wright$^{1}$ $^{2}$\thanks{Contact e-mail: \href{wright:a.j.wright@soton.ac.uk}{a.j.wright@soton.ac.uk}} and Ian Hawke$^{1}$ $^{2}$
\\
$^{1}$Mathematical Sciences and STAG Research Centre, University of Southampton, Southampton, SO17 1BJ, UK \\
$^{2}$Next Generation Computational Modelling Group, University of Southampton, Southampton, SO16 7PP, UK}

\date{Last updated 2015 May 22; in original form 2013 September 5}

\pubyear{2019}

\begin{document}
\label{firstpage}
\pagerange{\pageref{firstpage}--\pageref{lastpage}}
\maketitle

\begin{abstract}
We present an extension to the special relativistic, ideal magnetohydrodynamics (MHD) equations, designed to capture effects due to resistivity. The extension takes the simple form of an additional source term which, when implemented numerically, is shown to emulate the behaviour produced by a fully resistive MHD description for a range of initial data. The extension is developed from first principle arguments, and thus requires no fine tuning of parameters, meaning it can be applied to a wide range of dynamical systems. Furthermore, our extension does not suffer from the same stiffness issues arising in resistive MHD, and thus can be evolved quickly using explicit methods, with performance benefits of roughly an order of magnitude compared to current methods.
\end{abstract}





\section{Motivation}
    Modern simulations of astrophysical systems typically involve a number of physical models. For systems that exhibit strong gravitational fields, the equations of general relativity require evolution to describe how the geometry of spacetime develops. As the evolution of the spacetime is coupled to the matter within it, motion of the matter will change the geometry of spacetime, and the geometry of spacetime alters the motion of the matter.
	
	In totality, this system represents a highly non-linear, general relativistic, magneto-hydrodynamics (GRMHD) model. By-and-large, the equations of GRMHD conform to a type of MHD known as the ideal approximation \citep{Font2008}. For a wide range of systems, the ideal description of GRMHD performs well, and the simplification of perfect electrical conduction, among others, is generally accurate.
	
	There does exist, however, a number of problems in astrophysics that require a description of MHD with non-zero resistivity \citep{Palenzuela2009, Dionysopoulou2013, Andersson2016}. Simulations of binary neutron star mergers \citep*{Dionysopoulou2015}, magneto-rotational instabilities \citep{Qian2016}, magnetic reconnection \citep{Lyubarsky2005, Mignone2012} and jet launching from black-hole accretion \citep{Qian2018} have all shown a sensitive dependence upon the magnitude of the electrical resistivity. And yet, the number of large-scale numerical astrophysics codes that implement resistive GRMHD remains small.
	
	One issue with more physically complex models that prevents the wide spread adoption of the resistive GRMHD equations is the additional computation required in their evolution. Due to the possibly stiff source term of resistive GRMHD \citep{Palenzuela2009}, an implicit integrator, often the IMEX schemes of \citet{Pareschi2004}, is required to keep execution times reasonable \citep*{Palenzuela2009, Dionysopoulou2013, Aloy2016,  Miranda-Aranguren2017, Wright2019}. Whilst these schemes allow for a faster evolution than would be possible using traditional explicit schemes for the majority of the parameter space, they still result in at least a factor $5\times$ slow-down compared to conventional, ideal GRMHD models \citep{Dionysopoulou2015}. Furthermore, rewriting a well-tested code-base to evolve the resistive GRMHD equations is no small feat, requiring an increase in the number of fields to evolve, new flux and source vectors and new primitive recovery procedures, along with implementing and testing a new class of integrator. 
	
	\subsubsection*{Model extensions}
	
		Sub-grid source terms are beginning to see a number of applications in modelling astrophysical systems. The general principle behind these extensions is to include an additional source term into the equations of motion of the system, aimed at emulating the effects of unresolved fluid motion. 
		
		A common application of sub-grid sources is in the modelling of classical turbulence simulations, also known as large-eddy simulations (LES). In LES, the equations of motion are explicitly redefined in terms of resolved and unresolved quantities, and a closure is then assumed that relates the values of the sub-grid scale fields to those that are resolved in the simulation. Using this technique, it is possible to approximate the behaviour that would result from more finely resolved grids. \citet{Radice2017a} applied the classical Smagorinsky closure to the equations of general relativistic hydrodynamics for a merger simulation, showing that by modelling the sub-grid scale turbulence, the collapse of the hyper-massive neutron star remnant is altered. Similar work by \citet{Vigano2019} uses the LES closure approach to model unresolved motion of the non-relativistic MHD equations, with hopes of an extension to the full GRMHD system.
		
		In other work, \citet{Giacomazzo2015} developed a sub-grid model, similar to \citet{Palenzuela2015}, that captured how the behaviour of the magnetic fields is altered by unresolved motion. The motivation for such a source term comes from the assumption that the unresolved turbulent motion resulting from the Kelvin-Helmholtz instability (KHI) amplifies the magnetic fields. While this assumption may be justified \citep{Price2006, Obergaulinger2010, Kiuchi2015}, the free parameters in this model are constrained via local simulations of KHIs \citep{Zrake2013a}, rather than a first principles argument. 
		
		The benefits of these model extensions lie in the huge reduction in computational cost required to evolve the system. These extensions capture physics that occurs in unresolved regions of the domain, regions which would require a prohibitively high resolution to capture with conventional methods---the smallest scales on which the KHI develops during mergers simulations, for example, is on the order $\sim 10$\,cm \citep{Palenzuela2015}. Employing these methods allows one to capture the range of relevant physics in a fraction of the time, and with very little change to currently existing numerical codes.
		
		In this paper, we develop an extension to the special relativistic, ideal MHD equations, that captures the resistive effects present in the full, resistive MHD model. This extension, dubbed a resistive extension generated for ideal magneto-hydrodynamics (REGIME), is derived from first principles arguments, and as such requires no fine tuning of parameters for different astrophysical scenarios. 

		The rest of the paper is laid out as follows. In Section \ref{sec:MHD} we briefly present the two models of MHD that we are concerned with: special relativistic, ideal and resistive MHD. Section \ref{sec:CE expansion} derives the proposed source, details how one should implement this new extension numerically, and discusses its effect on the stability of simulations. Results of the extension are presented in Section \ref{sec:results}, and finally, in Section \ref{sec:discussion}, we summarise the findings of the previous sections, discuss how they fit into current astrophysical simulations, and propose the future direction of the project.

\section{Magneto-hydrodynamic Models}\label{sec:MHD}
	
	In this section, we outline two models of MHD that are used in relativistic astrophysics. In order to simplify the numerics in later sections and to test the validity of the method, we will limit ourselves to special relativity. In moving to a general relativistic description, only the form of the equations should change, and so the analysis we perform here should still apply---this includes implementations of GRMHD which evolve the vector potential instead of the magnetic fields \citep{Etienne2010}. We will also adopt the Einstein summation convention over repeated indices, where sums are over the three spatial dimensions, $\{x, y, z\}$, and $\delta_{ij}$ and $\epsilon_{ijk}$ are the Kronecker delta and Levi-Civita tensor, respectively. We use units where $\mu_0 = \epsilon_0 = 1$, \edit{such that $c=1$}, throughout.
	 
	 \subsection{Ideal MHD}
		The first model we present is that of ideal MHD, and this is the model that we will extend by means of the sub-grid source. Ideal MHD is the form most commonly used, for example, in merger simulations, due to the relative simplicity of its mathematical form, and the numerical methods employed in solving it. For a more comprehensive description of ideal MHD see \citet{Nobel2006, Font2008, Anton2010, Siegel2018}. We use the notation $\partial_t$ and $\partial_i$ to represent the partial derivative with respect to time and the $i^{\text{th}}$-spatial coordinate.
		
		The equations of motion, in conservative form, are given as
		
		\begin{align} \label{eq:ideal conservation}
			\partial_t
			\begin{pmatrix}
				D \\ S^j \\ \tau \\ B^k
			\end{pmatrix} + \partial_i
			\begin{pmatrix}
				Dv^i \\ S^jv^i + p^* \delta^{ij} - b^jB^i/W \\ \tau v^i + p^*v^i - b^0 B^i / W \\ v^iB^k - v^kB^i
			\end{pmatrix} = 
			0,
		\end{align}
		where the conserved quantities, $\{D, S^j, \tau, B^k\}$, correspond to the \edit{fluid} density, specific momentum in the $j^{\text{th}}$-direction, kinetic energy density and magnetic fields, and are related to the primitive quantities, $\{\rho, v^j, p, B^k\}$, namely the \edit{rest-mass} density, fluid velocity in the lab frame, hydrodynamic pressure and the magnetic field respectively, via
		\begin{equation}
			\begin{aligned}
				D &= \rho W, \\
				S^j &= \rho h^* W^2 v^j - b^0 b^j, \\
				\tau &= \rho h^* W^2 - p^* - (b^0)^2 - D.
			\end{aligned}
		\end{equation}
		In addition, we have the relations for the Lorentz factor, $W$, total pressure and specific enthalpy (i.e.\ including contributions from the magnetic fields), $p^{*}$ and $h^{*}$, and the relations of the components of the magnetic four-vector, $b^{\mu}$, to the lab-frame magnetic field, $B^i$:
		\begin{equation}
			\begin{aligned}
				W &= 1/\sqrt{1 - v_i v^i}, \\
				p^* &= p + b^2/2, \\
				h^* &= h + b^2 / \rho, \\
				b^0 &= W B_i v^i, \\
				b^i &= B^i/W + b^0v^i, \\
				b^2 &= B_i B^i/W^2 + (B_i v^i)^2.
			\end{aligned}
		\end{equation}
		The primitive quantities along with an equation of state completely define the system. Throughout this paper, we will assume a $\Gamma$-law equation of state, where $p=\rho e (\Gamma - 1)$, $e$ is the specific internal energy and $\Gamma$ is the ratio of specific heats, with the specific enthalpy defined as $h=1+e\Gamma$. Whilst the choice of the equation of state will change the form of the source term presented at the end of Section \ref{sec:numerics}, the general form of the source term will be unaltered.
		
		Although the system is defined by the primitive quantities, it is the conserved quantities that are evolved explicitly. This means, therefore, that there must be some scheme for transforming from the conserved variables to the primitives such that the fluxes in equation (\ref{eq:ideal conservation}) can be computed. The transformations can be found in detail in the aforementioned references.

	\subsection{Resistive MHD}
		The second model we present is resistive MHD, and is the model whose behaviour we wish to replicate by means of an extension of ideal MHD. Once again, further details of resistive MHD may be found in \citet{Palenzuela2009, Dionysopoulou2013}. The form of resistive MHD in balance law form reads
		\begin{align} \label{eq:resistive conservation}
			\partial_t 
			\begin{pmatrix}
				D \\ S_j \\ \tau \\ B^k \\ E^k \\ \varrho
			\end{pmatrix}
			+ \partial_i
			\begin{pmatrix}
				Dv^i \\ S^i_j \\ S^i - D v^i \\ \epsilon^{ijk} E_j\\ -\epsilon^{ijk} B_j \\ J^i
			\end{pmatrix}
			=
			\begin{pmatrix}
				0 \\ 0 \\ 0 \\ 0\\ -J^k \\ 0
			\end{pmatrix}.
		\end{align}
		The state vector has picked up four additional fields, three from the electric field and one corresponding to the charge density, $\{E^i, \varrho\}$ respectively. It is not essential that we evolve the charge density as one can compute it from Gauss' law, but we choose to do this to avoid difficulties when taking derivatives of near-discontinuous data.
		  
		Once again, we relate the conserved vector, the charge density current, $J_i$, and the momentum flux, $S_{ij}$, to the primitive quantities with,
		
		\begin{align} \label{eq:resistive prims}
			\begin{pmatrix}
				D \\ S_i \\ \tau \\ J_i \\ S_{ij} \\
			\end{pmatrix}
			=
			\begin{pmatrix}
				\rho W \\ \rho h W^2 v_i + \epsilon_{ijk}E^j B^k \\ \rho h W^2 - p + \frac{1}{2}(E^2 + B^2) - \rho W \\ \varrho v_i + W \sigma [E_i + \epsilon_{ijk}v^j B^k - (v_k E^k)v_i] \\ \rho h W^2 v_i v_j + [p + \frac{1}{2}(E^2 + B^2)] \delta_{ij} - E_i E_j - B_i B_j
			\end{pmatrix},
		\end{align}
		where all quantities are interpreted in the same way as for ideal MHD. In addition to the equation of state, the form of the charge current density is also needed to relate the electric and magnetic fields to the charge density and close the system. We also note the inclusion of the conductivity, $\sigma$, in the definition of the charge current density. This is to be expected, as for resistive MHD there is no assumption that charge flows perfectly (as there is in ideal MHD) and so the conductivity can take any finite, non-negative value. We shall see how this causes difficulties in the next section.
		
		We should also note that the conductivity we present here is scalar, but that in general can be described by a tensor \citep{Dionysopoulou2015}. While including the full tensor description would introduce additional terms into the source of the electric fields, it would not change the analysis that will follow. In this paper, we limit our analysis to a scalar conductivity commonly used in astrophysical settings \citep{Komissarov2008, Dionysopoulou2015, Mohseni2015, Miranda-Aranguren2017}.
		
		As is the case for ideal MHD, we also require some algorithm to compute the primitive from the conserved quantities, two examples of which are found within \citet{Palenzuela2009, Dionysopoulou2013}.

		\subsubsection*{Numerical difficulties}
		
			The major difference between the two models of MHD that are presented here lie in the form of the source terms. We can see that the source of the electric fields in the resistive model, equation (\ref{eq:resistive conservation}), is non-zero and proportional to the conductivity. As mentioned, charge flows without resistance in ideal MHD, and thus the fluid has an infinite conductivity. Clearly then, near the ideal limit when $\sigma \to \infty$, the source term in the resistive MHD equations begins to dominate. This is a common feature of balance laws equations, and can be interpreted as some driving term that acts on a much shorter timescale than the conservation system. 
			
			When the timescale that the source acts on is shorter than the timescale of the simulation, $\epsilon \lesssim \Delta t$, the system is said to be \textit{stiff}, \edit{where in this case $\epsilon \propto 1/\sigma$}. This results in some difficulties regarding the numerical evolution of the equations. In order to maintain a stable evolution, it is sufficient to either dramatically reduce the size of the timestep used in the simulation, or to employ a set of implicit or semi-implicit time integrators \citep{Pareschi2004}. Unfortunately, both of these approaches will increase the execution time of a simulation by around a factor of five or more, depending upon the magnitude of the conductivity.
			
			What would be useful is some source term that captures the behaviour due to finite conductivity, but that avoids the numerical difficulties in the ideal limit of the full, resistive MHD system. The following section will derive such a source term using a Chapman-Enskog-type analysis, but first we introduce the notation that will be used.
			
			We can re-write system (\ref{eq:resistive conservation}) in the following, more compact way:
			\begin{subequations}  \label{eq:cons law}
				\begin{align}
					\partial_t \bm{q} + \partial_i \bm{f^i}(\bm{q}, \overline{\bm{q}})&=\bm{s}(\bm{q}, \overline{\bm{q}}) \label{eq:cons law q}, \\
					\partial_t \bm{\overline{q}} + \partial_i \overline{\bm{f}}^i(\bm{q}, \overline{\bm{q}})&= \frac{\overline{\bm{s}}(\bm{q}, \overline{\bm{q}})}{\epsilon}, \label{eq:cons law qbar}
				\end{align}
			\end{subequations}
			where we indicate equations which become stiff as $\sigma \to \infty$ with an over-bar. This means that $\overline{\bm{q}} = \{E_x, E_y, E_z\}$ with the corresponding fluxes, $\overline{\bm{f}^i}(\bm{q}, \overline{\bm{q}})$, and sources, $\overline{\bm{s}}(\bm{q}, \overline{\bm{q}})$, taken from equation (\ref{eq:resistive conservation}). The remaining variables are non-stiff in the ideal limit, and denoted $\bm{q} = \{D, S_x, S_y, S_z, \tau, B_x, B_y, B_z\}$, where we have left out the charge density. We will also denote the vector of primitive variables with $\bm{w} = \{\rho, v_x, v_y, v_z, p, B_x, B_y, B_z, E_x, E_y, E_z, \varrho\}$, where $\varrho$ is the charge density.
			
			As an aside, one might notice that the flux in the charge density evolution may be of the same size as the source of the electric fields, and so the equation can also be stiff. Instead, if we were to multiply this equation by the small timescale, $\epsilon$, we can make the equation non-stiff. Regardless, as the charge density is, in fact, defined by Gauss's law, the choice of evolving it is purely a numerical one and so it can be excluded from the following expansion.
	
\section{Chapman-Enskog Expansion}\label{sec:CE expansion}
	
	In this section we will use the Chapman-Enskog (CE) method of expansion to derive the form of the REGIME source term. The CE expansion was initially used to determine higher order moments for solutions to the Boltzmann equation \citep{Denicol2018}. By expressing the particle distribution function in terms of powers of some small quantity, one can approximate the true solution up to arbitrary order, with each subsequent power encapsulating some new physical phenomenon. 
	
	An additional application of this expansion is presented in \citet{LeVeque2002}, in which LeVeque demonstrates how a coupled system of potentially stiff balance law equations may be reduced, in a given limit, to a single, modified system. Here, we apply the same logic to ideal and resistive MHD. We consider the solution to ideal MHD to be the equilibrium system of resistive MHD, and derive the form of a perturbation on top of this. As we will see, this perturbation looks like a diffusion term, as is the case in LeVeque's analysis.
	
	\subsection{System Derivation}
		We begin with equations (\ref{eq:cons law}), recalling that the non-stiff and stiff conserved variables are labelled $\bm{q}$ and $\overline{\bm{q}}$ respectively. In order to maintain finite solutions in the ideal limit, we require that $\lim_{\epsilon \to 0} \overline{\bm{s}}(\bm{q}, \overline{\bm{q}})=\bm{0}$. We will refer to the solution in the limit $\epsilon \to 0$ as the equilibrium system, as for the models we are interested in here this corresponds to the RHS $= 0$, and equation (\ref{eq:cons law q}) reduces to ideal MHD. In the case of resistive MHD, this limit corresponds to ideal MHD where $\sigma = \infty$. Therefore, for large but finite conductivities, we can think of the solution of resistive MHD as some perturbation on top of the solution of ideal MHD. \edit{This corresponds to small $\epsilon$ when compared to other timescales in the simulation, for example $\Delta t$.}
		
		To derive the form of this perturbation, we need to define the equilibrium solution. At this point, we assume that the stiff variables have some values given by $\overline{\bm{q}}_0$, and so the stiff source in the equilibrium limit must be 
		\begin{align}\label{eq:zero source}
			\overline{\bm{s}}(\bm{q}, \overline{\bm{q}}_0) = \bm{0}.
		\end{align}
		
		The Chapman-Enskog expansion involves expressing the stiff variables in terms of a series of increasing powers of the small quantity, $\epsilon$. Expanding the solution for the stiff variables about their equilibrium solution gives
		\begin{align}
			\overline{\bm{q}} = \overline{\bm{q}}_0 + \epsilon \overline{\bm{q}}_1 + \epsilon^2 \overline{\bm{q}}_2 + \mathcal{O}(\epsilon^3),
		\end{align}
		in which each new power of $\epsilon$ represents some higher order perturbation on top of the previous. Using the Taylor expansion of some vector function as 
		\begin{equation*}
    		\begin{aligned}
        		\bm{A}(\overline{\bm{q}}_0 + \epsilon \overline{\bm{q}}_1 + \mathcal{O}(\epsilon^2)) = \bm{A}(\overline{\bm{q}}_0) + \epsilon \partial_{\overline{\bm{q}}}\bm{A}(\overline{\bm{q}})|_{\overline{\bm{q}}=\overline{\bm{q}}_0} \cdot \overline{\bm{q}}_1 +  \mathcal{O}(\epsilon^2),
    		\end{aligned}
		\end{equation*}
		we can expand the stiff and non-stiff variables, equation (\ref{eq:cons law}), about the equilibrium solution. Note, we are considering a 1D system for simplicity---in 3D the following analysis is qualitatively the same, and will be presented in Section \ref{sec:higher D}.
		
		Keeping only leading order terms, we get an evolution equation for the equilibrium state of the stiff system up to order $\mathcal{O}(\epsilon^0)$:
		\begin{align} \label{eq:reduced qbar cons}
			\partial_t{\overline{\bm{q}}_0}  + \partial_x  \overline{\bm{f}}_0 = \pdv{\overline{\bm{s}}_0}{\overline{\bm{q}}} \overline{\bm{q}}_1.
		\end{align}
		We will use the notation $\bm{A}_0$ to represent the vector $\bm{A}(\bm{q}, \overline{\bm{q}}_0)$, evaluated at the equilibrium point---therefore, $\partial{\bm{s}(\bm{q}, \overline{\bm{q}})}/{\partial \overline{\bm{q}}}|_{\overline{\bm{q}}=\overline{\bm{q}}_0} \equiv \partial \overline{\bm{s}}_0 / \partial \overline{\bm{q}}$.
		
		We now wish to determine the form for the first order perturbation of the stiff variables, $\overline{\bm{q}}_1$. First, we assume that the equilibrium state of the stiff variables can be fully characterised by the non-stiff variables---that is, $\overline{\bm{q}}_0 = \overline{\bm{q}}_0(\bm{q})$. For the case of resistive and ideal MHD, we know this is true as the electric fields can be expressed as $\bm{E} = - \bm{v} \times \bm{B}$. Therefore, as
		\begin{align*}
			\pdv{\overline{\bm{q}}_0}{t} = \pdv{\overline{\bm{q}}_0}{\bm{q}} \pdv{\bm{q}}{t},
		\end{align*}
		and the evolution of the non-stiff variables, up to zeroth order in $\epsilon$, is
		\begin{align} \label{eq: reduced cons law q}
			\partial_t \bm{q} + \partial_x \bm{f}_0 = \bm{s}_0,
		\end{align}
		we can remove the time dependence of the first order perturbation, equation (\ref{eq:reduced qbar cons}), and re-arrange for $\overline{\bm{q}}_1$:
		\begin{align}
			\overline{\bm{q}}_1 &= \bigg( \frac{\partial \overline{\bm{s}}_0}{\partial \overline{\bm{q}}} \bigg)^{-1} \Bigg[ \frac{\partial \overline{\bm{q}}_0}{\partial \bm{q}} \bigg( \bm{s}_0 - \frac{\partial \bm{f}_0}{\partial x} \bigg) + \frac{\partial \overline{\bm{f}}_0}{\partial x} \Bigg].  \label{eq: general form correction}
		\end{align}
		Here, we note that the non-stiff flux evaluated at the equilibrium point is only a function of the non-stiff variables, $\bm{f}(\bm{q}, \overline{\bm{q}}_0(\bm{q})) \equiv \bm{f}(\bm{q}) \equiv \bm{f}_0 \equiv \bm{f}$.
		
		Upon substituting the form for the perturbation, equation (\ref{eq: general form correction}), into the expanded evolution equation for the non-stiff variables, and keeping orders up to $\mathcal{O}(\epsilon)$, we can rewrite the perturbed system in an intuitive and compact form:
		\begin{align}\label{eq:new nonstiff no epsilon}
		\pdv{\bm{q}}{t} + \pdv{(\bm{f} + \hat{\bm{f}})}{x} = \bm{s}_0 + \hat{\bm{s}} + \pdv{\hat{\bm{D}}}{x}.
		\end{align}
		where we have absorbed the timescale, $\epsilon$, into the definition of the stiff source, $\bar{\bm{s}}$. 
		
		In this form, the modified flux term, modified source term, and the diffusion-like term are
		\begin{subequations}\label{eq:modifications}
			\begin{align}
				\hat{\bm{f}} &= \frac{\partial \bm{f}}{\partial \overline{\bm{q}}}  \bigg( \frac{\partial \overline{\bm{s}}_0}{\partial \overline{\bm{q}}} \bigg)^{-1} \frac{\partial \overline{\bm{q}}_0}{\partial \bm{q}} \bm{s}_0 \\
				\hat{\bm{s}} &= \frac{\partial \bm{s}_0}{\partial \overline{\bm{q}}} \bigg( \frac{\partial \overline{\bm{s}}_0}{\partial \overline{\bm{q}}} \bigg)^{-1} \bigg[ \bigg( \frac{\partial \overline{\bm{f}}_0}{\partial x} - \frac{\partial \overline{\bm{q}}_0}{\partial \bm{q}} \frac{\partial \bm{f}}{\partial x} \bigg)+ \frac{\partial \overline{\bm{q}}_0}{\partial \bm{q}} \bm{s}_0 \bigg] \\
				\hat{\bm{D}} &= -\frac{\partial \bm{f}}{\partial \overline{\bm{q}}}  \bigg( \frac{\partial \overline{\bm{s}}_0}{\partial \overline{\bm{q}}} \bigg)^{-1} \bigg[ \frac{\partial \overline{\bm{f}}_0}{\partial x} - \frac{\partial \overline{\bm{q}}_0}{\partial \bm{q}} \frac{\partial \bm{f}}{\partial x} \bigg]. \label{eq:diffusion vector}
			\end{align}
		\end{subequations}
		The hats in equations (\ref{eq:modifications}) signify that these terms are perturbations on top of the equilibrium system, coming in at order $\mathcal{O}(\epsilon^1)$. 
		
		Looking at the form for the ideal MHD equations (\ref{eq:ideal conservation}), we can see that there is zero source term---$\bm{s}(\bm{q}) \equiv 0$. As a result, the modified flux and source terms in the previous equations reduce to zero, and we are left with only the diffusion term being non-zero. This means the equations of motion of ideal MHD are modified only by a single diffusion term\footnote{We note that general relativistic, ideal MHD has a non-zero source term, and so this simplification may not be made.},
		\begin{align}\label{eq:new nonstiff}
			\pdv{\bm{q}}{t} + \pdv{\bm{f}}{x} = \pdv{\hat{\bm{D}}}{x}.
		\end{align}
		
		We can see that in the ideal limit, which corresponds to $\{\sigma, \epsilon\} \to \{ \infty, 0\}$, the fact that $\hat{\bm{D}} \propto \epsilon$ means equation (\ref{eq:new nonstiff}) reduces to the standard form for ideal MHD, as expected. Then, as the conductivity reduces, larger corrections are made through the diffusive source term to mimic behaviour that should be present in resistive MHD.
		
		Finally, observe how the source term for resistive MHD is proportional to $\sigma$, but that the REGIME source term scales as $\epsilon \propto \sigma^{-1}$. This means that the two models become stiff in opposing limits---near the ideal regime (large $\sigma$) REGIME will be stable as a result of a small source term, and will only become stiff, and potentially unstable, as $\sigma \to 0$. The big benefit of this behaviour is that near the ideal regime we can confidently evolve REGIME with explicit time integrators, knowing that source contributions will remain small. In contrast, in the event of very low conductivities, $\sigma \sim 0$, it will not be sensible to evolve REGIME using implicit schemes, not least because in this regime resistive MHD is likely to be stable with explicit integrators. Because the numerical flux function appears in the diffusion vector (i.e.\ $\partial_x \bm{f}$), and this has a dependence on neighbouring cells, an implicit integration for REGIME would require solving the system for all cells in the domain at once (i.e.\ an $N_x \times N_y \times N_z \times N_{cons}$ dimensional root-find, where $N_{cons}$ is the size of the conserved vector).
		
	\subsection{Numerics} \label{sec:numerics}
	
		Now we have the form for the source term, our attention turns to implementing it numerically. The interpretation of a term in a fluid's equations of motion may be intuited by the order of the spatial derivative of the conserved fields. For example, advection is noted by a first order derivative, diffusion by a second order derivative and dispersion via a third order derivative. This is why we call the new piece in equation (\ref{eq:new nonstiff}) a diffusion-like term, as it can be re-written to include second order spatial derivatives of the conserved fields:
		\begin{align}\label{eq:second conserved derivative}
			\pdv{\bm{q}}{x} + \pdv{\bm{f}}{x} = -\pdv{}{x} \bigg[ \frac{\partial \bm{f}}{\partial \overline{\bm{q}}}  \bigg( \frac{\partial \overline{\bm{s}}_0}{\partial \overline{\bm{q}}} \bigg)^{-1} \bigg( \frac{\partial \overline{\bm{f}}_0}{\partial \bm{q}} - \frac{\partial \overline{\bm{q}}_0}{\partial \bm{q}} \frac{\partial \bm{f}}{\partial \bm{q}} \bigg) \pdv{\bm{q}}{x} \bigg],
		\end{align}
		recalling that both $\hat{\bm{f}} = \hat{\bm{s}} = \bm{0}$ when considering ideal and resistive MHD in the special relativistic limit.
		
		In order to implement this new piece, it is useful to understand more about what it involves. If we look that the diffusion vector given in equation (\ref{eq:diffusion vector}), the first term is the Jacobian of the non-stiff system with respect to the stiff variables, $\partial_{\overline{\bm{q}}}\bm{f}$, and is therefore a matrix operation---as, indeed, is the second and fourth term. The third and fifth terms correspond to the spatial derivatives of the stiff and non-stiff flux vectors. In fact, the form of the diffusion term in equation (\ref{eq:diffusion vector}) is preferable to that in equation (\ref{eq:second conserved derivative}), as any numerical flux function that computes $\mathcal{\bm{F}}(\bm{q}) = \partial_x \bm{f}(\bm{q})$ can be re-used \textit{as is} for the generation of the source term.
		
		All that is required then is to determine the forms of the two matrices,
		\begin{subequations}
			\begin{align}
				\underline{\underline{M}}_1 &= \frac{\partial \bm{f}}{\partial \overline{\bm{q}}}  \bigg( \frac{\partial \overline{\bm{s}}_0}{\partial \overline{\bm{q}}} \bigg)^{-1}, \\
				\underline{\underline{M}}_2 &= \frac{\partial \overline{\bm{q}}_0}{\partial \bm{q}}.
			\end{align}
		\end{subequations}
		
		Differentiating with respect to the conserved variables is challenging, as we do not know how to express the flux and source vectors in terms of only the conserved quantities. We can, however, express the conserved, flux and source vectors in terms of the primitive variables, and thus we can differentiate using the primitive quantities. For example,
		\begin{align}
			\pdv{\bm{f}}{\bm{q}} = \pdv{\bm{f}}{\bm{w}} \pdv{\bm{w}}{\bm{q}},
		\end{align}
		with $\bm{w}$ as the vector of primitive variables. In this way, the new form for the $M$ matrices is
		\begin{subequations} \label{eq:invert matrices}
			\begin{align}
				\underline{\underline{M}}_1 &= \pdv{\bm{f}}{\bm{w}}  \pdv{\bm{w}}{\overline{\bm{q}}} \bigg( \pdv{\overline{\bm{s}}_0}{\bm{w}}  \pdv{\bm{w}}{\overline{\bm{q}}} \bigg)^{-1} \\
				&= \pdv{\bm{f}}{\bm{w}}   \bigg( \pdv{\overline{\bm{s}}_0}{\bm{w}} \bigg)^{+} \\
				\underline{\underline{M}}_2 &= \pdv{\overline{\bm{q}}_0}{\bm{w}} \bigg( \pdv{\bm{q}}{\bm{w}} \bigg)^+,
			\end{align}
		\end{subequations}
		
		Here, we have used the superscript $^+$ to denote the Moore-Penrose pseudoinverse \citep{Barata2011}. As the length of the vectors $\overline{\bm{s}}_0$ and $\bm{w}$ are, in general, not the same, the resulting matrix will not be square, nor have a corresponding inverse. As a result, we use the definition of the right-pseudoinverse of a matrix $Q$ as, $Q^+ = Q^T (Q Q^T)^{-1}$. We will use the term inverse to refer to the pseudoinverse henceforth. 

		Here, we have a choice of how to compute the matrices of interest---that is we can invert them numerically, or try to get the form of the inverted matrix symbolically. Inverting matrices numerically, especially when densely populated, can require a large amount of computation, reducing accuracy as well as slowing down simulations. If the algebraic form of the matrices were at hand, this would lead to a far more efficient simulation, and as we are trying to build a source term to extend ideal MHD with the intention of being faster to evolve than resistive MHD, it is sensible to adopt the performance gains of a purely symbolic source term. 
		
		In order to generate human readable terms for equations (\ref{eq:invert matrices}), we have had to make a number of assumptions. Firstly, we will assume a low velocity limit, in which terms of $\mathcal{O}(v^2)$ can be ignored. For instance, with this approximation in resistive MHD, the $y$-momentum flux in the $x$-direction reduces to $S_{12} = -E_xE_y - B_xB_y$. Secondly, to reduce the number of terms in the inverted matrices, we assume that the fluid only couples weakly to the magnetic field, enforcing this by setting the electric and magnetic fields to zero. As the electric and magnetic fields are still present in the un-inverted matrices, their influence still makes it through to the final source term. As a check of these assumptions and their effect on the system we also implemented the REGIME source term numerically with no approximations. After doing this, we found virtually no difference in the simulation output for mildly relativistic flows, $v \sim 0.5c$. Making these simplifications, we can use a symbolic algebra program to compute the form for the source term. For this task, we used Wolfram Mathematica v11.2 \citep{WolframResearch2018}. The full notebooks, along with greater detail, are available through the METHOD GitHub page\footnote{\edit{https://github.com/AlexJamesWright/METHOD}}, so we will only show the final forms here. 
		
		For the systems of equations presented here, namely ideal and resistive MHD in the special relativistic limit and with the above assumptions made, we get the simple result that $\underline{\underline{M}}_2 = 0$. The reason for this is that $\partial \overline{\bm{q}}_0/\partial \bm{w}$ is a $3\times12$ matrix in which only three entries are non-zero---these correspond to the $\partial E_x / \partial E_x$, $\partial E_y / \partial E_y$ and $\partial E_z / \partial E_z$ terms. These terms are then dotted with $\partial \bm{E}/\partial \bm{S}$, which on account of the weak coupling approximation is zero.
		
		With regards to the $\underline{\underline{M}}_1$ matrix, we get
		\begin{equation}
			\begin{aligned}
			\frac{\partial \bm{f}}{\partial \bm{w}}^T &= 
				\begin{bmatrix}
					v_x  & 0 & 0 & 0 & 0 & 0 & 0 & 0 \\
					\rho & 0 & 0 & 0 & \frac{\Gamma p}{\Gamma - 1} & 0 & 0 & 0 \\
					0 & 0 & 0 & 0 & 0 & 0 & 0 & 0  \\
					0 & 0 & 0 & 0 & 0 & 0 & 0 & 0  \\
					0 & 1 & 0 & 0 & \frac{\Gamma v_x}{\Gamma - 1} & 0 & 0 & 0 \\
					0 & -B_x & -B_y & -B_z & 0 & 0 & 0 & 0 \\
					0 & B_y & -B_x & 0 & -E_z & 0 & 0 & 0  \\
					0 & B_z & 0 &-B_0 & E_y & 0 & 0 & 0  \\
					0 & -E_x & -E_y & E_z & 0 & 0 & 0 & 0  \\
					0 & E_y & -E_x & 0 & B_z & 0 & 0 & 0  \\
					0 & E_z & 0 & -E_x & -B_y & 0 & -1 & 0  \\
					0 & 0 & 0 & 0 & 0 & 1 & 0 & 0 
				\end{bmatrix},
				\end{aligned}
				\end{equation}
				\begin{equation}
				\begin{aligned}
				\label{eq:dwdsb}
				\bigg(\frac{\partial \overline{\bm{s}}}{\partial \bm{w}}\bigg)^{+} &= \alpha
				\begin{bmatrix}
				0 & 0 & 0 \\
				\mathcal{A}_{11} & \mathcal{A}_{12} & \mathcal{A}_{13} \\
				\mathcal{A}_{21} & \mathcal{A}_{22} & \mathcal{A}_{23} \\
				\mathcal{A}_{31} & \mathcal{A}_{32} & \mathcal{A}_{33} \\
				0 & 0 & 0 \\
				\mathcal{B}_{11} & \mathcal{B}_{12} & \mathcal{B}_{13} \\
				\mathcal{B}_{21} & \mathcal{B}_{22} & \mathcal{B}_{23} \\
				\mathcal{B}_{31} & \mathcal{B}_{32} & \mathcal{B}_{33} \\
				\mathcal{C}_{11} & \mathcal{C}_{12} & \mathcal{C}_{13} \\
				\mathcal{C}_{21} & \mathcal{C}_{22} & \mathcal{C}_{23} \\
				\mathcal{C}_{31} & \mathcal{C}_{32} & \mathcal{C}_{33} \\
				\mathcal{D}_{1} & \mathcal{D}_{2} & \mathcal{D}_{3},
				\end{bmatrix},
			\end{aligned}
		\end{equation}
		
		where we define the elements of the matrices $\mathcal{A}, \mathcal{B}$ and $\mathcal{C}$ as
		\begin{align}
			\mathcal{A}_{ij} &= -\varrho \sigma^2 B_i B_j + \epsilon_{ijk} B_k (\varrho^2 + \sigma^2) \sigma - \delta_{ij} \varrho (\varrho^2 + \sigma^2), \\
			\mathcal{B}_{ij} &= B_j \sigma^3 \epsilon_{ikl}v_k B_l - \epsilon_{ijk} v_k (\varrho^2 + \sigma^2) \sigma, \\
			\mathcal{C}_{ij} &= -\sigma^3 B_i B_j - \delta_{ij} \sigma (\varrho^2 + \sigma^2),
		\end{align}
		the vector $\mathcal{D}$ as
		\begin{align}
			\mathcal{D}_i = -v_i (\varrho^2 + \sigma^2) - \sigma^2 B_i v_k B_k,
		\end{align}
		and the pre-factor, $\alpha$, as
		\begin{align}
			\alpha = \frac{1}{(\varrho^2 + \sigma^2)(\varrho^2 + (1 + B^2)\sigma^2)}.
		\end{align}
		
		The pre-factor, $\alpha$, in equation (\ref{eq:dwdsb}) now acts in a similar way to the previous timescale, $\epsilon$. That is, in the ideal limit, $\alpha \propto \sigma^{-4} \to 0$, and the source term tends to zero, recovering ideal MHD. For large but finite conductivities however, the source term will modify the solution.
		
		It should also be noted that in equation (\ref{eq:dwdsb}), the charge density, $\varrho$, appears a number of times. As this term is not explicitly evolved in ideal MHD, we compute its value from Gauss' law, $\bm{\nabla} \cdot \bm{E} = \varrho$. 
		\edit{Furthermore, as the electric fields are not evolved in the ideal prescription, we use the assumption of our previous expansion, equation (\ref{eq:zero source}). This condition, that $\overline{\bm{s}}(\bm{q}, \overline{\bm{q}}_0) = 0$, implies that there is zero current, $\bm{J}=0$. For this to be valid as $\sigma \to \infty$, it would require that the leading order part, proportional to $\sigma$, is zero, and thus that $\bm{E} = -\bm{v}\times \bm{B}$, which is simply the ideal form for the electric fields.}
		
		All that remains to compute the diffusion vector is the $\partial_x \overline{\bm{f}}_0$ term. This can be done using any existing numerical flux function, which in the code used for this paper is a WENO3 reconstruction \citep{Shu1997}. 
		
		When determining the derivative of the diffusion vector, there are a number of numerical methods available, each of which can have an affect on the accuracy of the solution and the overall stability. For example, employing MINMOD slope limiting to reduce the onset of Gibbs oscillations also improves the accuracy of the solution in most scenarios. Despite this, in Kelvin-Helmholtz simulations, Section \ref{sec:KHI}, limiting in this way leads to excessive growth of the magnetic fields, and attempts to identify alternative slope limiting methods have all resulted in similar behaviour. As a result, we simply employ second-order central differencing, which we find gives good results across all simulations and is generally stable at shocks.

	\subsection{Stability Criterion}\label{sec:stability theory}
		The source term we have derived is diffusion-like, containing a second order spatial derivative. Whilst the original equilibrium system is in a strongly hyperbolic form, the addition we have made would be parabolic on its own, i.e.\ if there were no flux. As a result, we must first understand the limits on the spatio-temporal resolution such that the new system is stable. 
		
		Recall our system has the following form:
		\begin{align}
			\pdv{\bm{q}}{t} + \pdv{\bm{f}}{x} &= \pdv{x} \bigg( - \underline{\underline{M}}_1 \cdot \pdv{\overline{\bm{f}}_0}{x} \bigg).
		\end{align}
		In order to proceed, we will make the assumption that in a small region of the domain, $\underline{\underline{M}}_1$ is approximately constant. We will also rewrite the stiff flux derivative using the Jacobian, and further assume that it, too, is constant in  this region. Finally, if we assume that the resolution constraints set by the hyperbolic system are met, and that we are only interested in the second derivative term, then we can simplify and say
		\begin{align} \label{eq:simplified diffusion 1D}
			\pdv{\bm{q}}{t} &= \underline{\underline{K}} \cdot \pdv{^2 \bm{q}}{x^2},
		\end{align}
		
		Here, we make an analogy with the scalar case of the diffusion equation, $\partial_t q = k \partial_x^2 q$, in which an explicit scheme is generally considered stable if
		\begin{align}
			k \frac{\Delta t}{\Delta x^2} \le \frac{1}{2},
		\end{align}
		in which $\Delta t$ and $\Delta x$ are the size of the timestep and spatial resolution in a simulation.
		
		For the system, we suggest that $k$ be given by the largest magnitude eigenvalue of $\dunderline{K}$. This, therefore, requires an eigenanalysis of the matrix 
		\begin{align}\label{eq:Kmatrix}
			\dunderline{K} = - \pdv{\bm{f}}{\bm{w}} \bigg( \pdv{\overline{\bm{s}}_0}{\bm{w}} \bigg)^{-1} \pdv{\overline{\bm{f}}_0}{\bm{w}} \bigg( \pdv{\bm{q}}{\bm{w}} \bigg)^{-1}.
		\end{align}
		
		With the help of Mathematica (these scripts can also be found on the GitHub page), it can be shown that the largest eigenvalue is given by
		\begin{align}
			\lambda_\text{max} = \frac{\sigma \big(\varrho + \sigma^2 [1 + B_y^2 + B_z^2]\big)}{(\varrho^2 + \sigma^2) \big(\varrho^2 + \sigma^2[1+B^2] \big)} \sim \sigma^{-1}.
		\end{align}
	
		This result suggests that, assuming a CFL constraint given by $\Delta t / \Delta x = \nu$, the spatial resolution of any simulation using this source term must (roughly) satisfy
		\begin{align}\label{eq:stability criterion}
			\Delta x > \frac{2 \delta \nu}{\sigma},
		\end{align}
		where $\delta$ is some softening factor which we determine in section \ref{sec:stability results}. Intuitively, this means there is a limit on how well a simulation may be resolved. For high conductivities, this limit is not an issue as the RHS tends to zero, allowing (almost) arbitrarily fine resolutions. For resistive simulations, however, there is a finest grid resolution for which the system is numerically stable (for a given Courant factor), as the source term begins to become large compared to the hyperbolic part. Of course, one may always reduce the CFL condition and take smaller timesteps in order to achieve higher resolutions.
		
	\subsection{Higher dimensions}\label{sec:higher D}
		
		This section has so far assumed a one-dimensional system to simplify the analysis of the proposed model. In this subsection, we relax that assumption. As the majority of the analysis carries over to higher dimensions in the same manner, we only present the bottom line results.
		
		In general, we will have the non-stiff and stiff systems defined via
		\begin{align}
			\pdv{ \bm{q}}{t} + \pdv{ \bm{f}^{i}}{x^{i}} &= \bm{s}, \\
		\pdv{ \overline{\bm{q}}}{t} + \pdv{ \overline{\bm{f}}^{i}}{x^{i}} &= \frac{\overline{\bm{s}}}{\epsilon}.
		\end{align}
		The index $i$ spans the three spatial dimensions, such that $\bm{f}^{i}$ is the non-stiff flux vector in the $i^{th}$ coordinate direction, and $\partial/\partial x^{i}$ is the derivative with respect to the $i^{th}$ coordinate direction.

		\begin{figure*}
			\includegraphics[width=\linewidth]{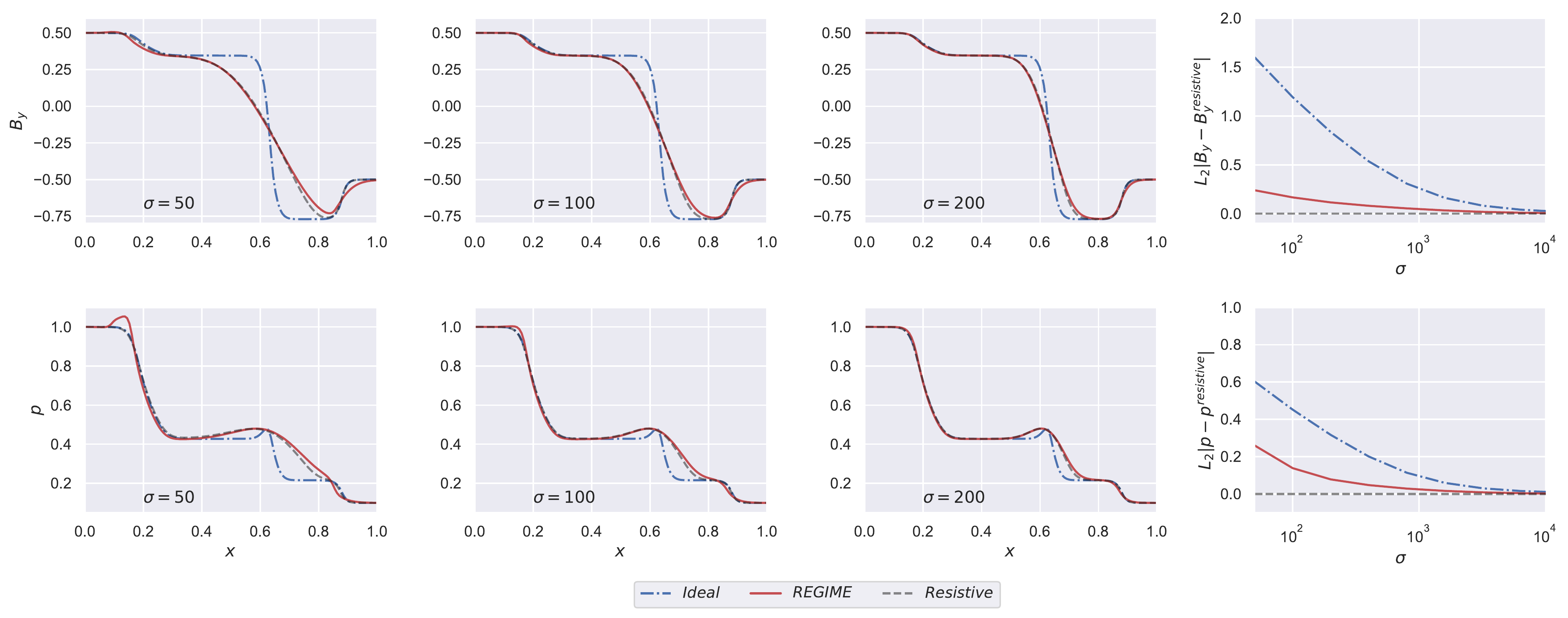}
			\caption{First three columns: Final state of the Brio-Wu test problem for a range of conductivities. Top row is the magnetic field in the $y$-direction, bottom row is the hydrodynamic pressure, each for the highlighted conductivity, $\sigma$. Rightmost column: The $L_2$ error-norm for the Brio-Wu test of ideal MHD and REGIME, using resistive MHD as the \textit{exact} solution. The error of both models grows as the system becomes more resistive, however REGIME's error grows many factors more slowly.}
			\label{fig:BrioWu final and error}
		\end{figure*}
		
		Applying the same expansion about the equilibrium system, the perturbed system can be written, in analogy with equation (\ref{eq:new nonstiff no epsilon}) such that the small scale is absorbed into the definition of $\hat{\bm{f}}, \hat{\bm{s}}$ and $\hat{\bm{D}}$, as
		\begin{align}
			\pdv{\bm{q}}{t} + \pdv{(\bm{f}^{i} + \hat{\bm{f}}^{i} )}{x^{i}} = \bm{s}_0 + \hat{\bm{s}} + \pdv{ \hat{\bm{D}}^{i}}{x^{i}}.
		\end{align}
		We note here that each spatial dimension has a corresponding modified flux and diffusion vector. The definitions of all modifications are given by the following expressions:
		\begin{subequations}\label{eq:modifications multiD}
			\begin{align}
			\hat{\bm{f}}^{i} &= \frac{\partial \bm{f}^{i}}{\partial \overline{\bm{q}}}  \bigg( \frac{\partial \overline{\bm{s}}_0}{\partial \overline{\bm{q}}} \bigg)^{-1} \frac{\partial \overline{\bm{q}}_0}{\partial \bm{q}} \bm{s}_0, \\
			\hat{\bm{s}} &= \frac{\partial \bm{s}_0}{\partial \overline{\bm{q}}} \bigg( \frac{\partial \overline{\bm{s}}_0}{\partial \overline{\bm{q}}} \bigg)^{-1} \bigg[ \bigg( \frac{\partial \overline{\bm{f}}_0^{j}}{\partial x^{j}} - \frac{\partial \overline{\bm{q}}_0}{\partial \bm{q}} \frac{\partial \bm{f}^{j}}{\partial x^{j}} \bigg)+ \frac{\partial \overline{\bm{q}}_0}{\partial \bm{q}} \bm{s}_0 \bigg], \\
			\hat{\bm{D}}^{i} &= -\frac{\partial \bm{f}^{i}}{\partial \overline{\bm{q}}}  \bigg( \frac{\partial \overline{\bm{s}}_0}{\partial \overline{\bm{q}}} \bigg)^{-1} \bigg[ \frac{\partial \overline{\bm{f}}_0^{j}}{\partial x^{j}} - \frac{\partial \overline{\bm{q}}_0}{\partial \bm{q}} \frac{\partial \bm{f}^{j}}{\partial x^{j}} \bigg],
			\end{align}
		\end{subequations}
		where we recall that repeated indices indicate a summation. Clearly, as we move to higher dimensions, the amount of computation required increases correspondingly---3D simulations require the calculation of three diffusion terms, compared to only one for a 1D simulation. We discuss the performance of REGIME in section \ref{sec:performance}. 
		
		With regards to the stability analysis, equation (\ref{eq:simplified diffusion 1D}) becomes,
		\begin{align}
			\pdv{\bm{q}}{t} = \underline{\underline{K}}^{ij} \cdot \pdv{^2 \bm{q}}{x^{i} \partial x^{j}},
		\end{align}
		where there are now nine matrices,
		\begin{align}
			\underline{\underline{K}}^{{ij}} = - \pdv{\bm{f}^{i}}{\bm{w}} \bigg( \pdv{\overline{\bm{s}}}{\bm{w}} \bigg)^{-1} \pdv{\overline{\bm{f}}^{j}}{\bm{w}} \bigg( \pdv{\bm{q}}{\bm{w}} \bigg)^{-1}.
		\end{align}
		There are now cross terms in the derivative, but to simplify the analysis we once again take the largest eigenvalue of all $K^{ij}$ to be the value of $k$ as in equation (\ref{eq:Kmatrix}). The largest eigenvalue is the same as in the previous section, and so we assume the same form for the resolution criterion, equation (\ref{eq:stability criterion}). As this is an order of magnitude estimation of the stability, we assume that any effects due to the additional terms are captured in $\delta$ to be determined in Section \ref{sec:stability results}.

\section{Results}\label{sec:results}

	In this section we present a series of results relating to the convergence of REGIME with the conductivity, the stability of the proposed model and its performance relative to the resistive model. We then end this section with a discussion on the current limitations of this model.
	
	\begin{figure*}
		\centering
		\includegraphics[width=\linewidth]{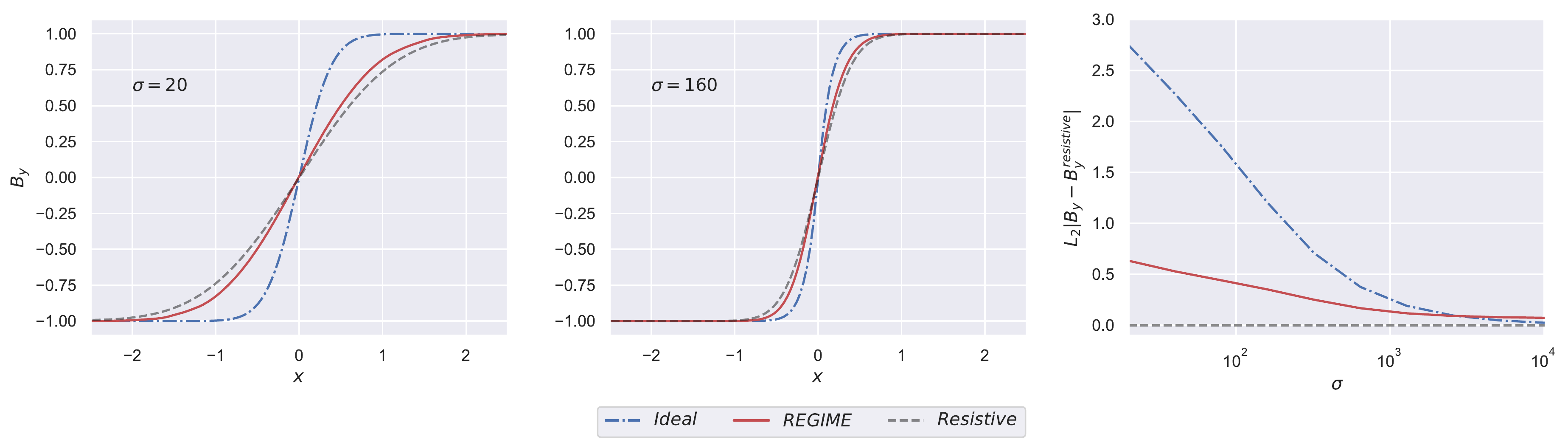}
		\caption{Left and middle: Final state comparison of the $y$-direction magnetic field for the self-similar current sheet problem. Larger resistivities lead to a greater rate of magnetic diffusion and thus greater smearing out compared to the ideal MHD solution. Right: Error growth for the self-similar current sheet problem. The error is calculated as the $L_2$-norm of the difference of the $y$-direction magnetic fields, using resistive MHD as the \textit{exact} solution. The difference between resistive MHD and REGIME is small down to $\sigma=20$, whereas ideal MHD quickly deviates from the exact solution due to the lack of any resistive diffusion. The REGIME source term captures this diffusion well, even for very small conductivities. \edit{The difference in the solutions of ideal MHD are due to differing initial data as a result of different $\sigma$.}}
		\label{fig:CurrentSheet final and error}
	\end{figure*}
	
	For the convergence plots presented here, we will consider the results of the resistive MHD model to be the \textit{exact} solution. Furthermore, any performance results have been optimised to be the fastest execution for a given set of parameters---for example, the runtimes represent the largest timestep (optimum CFL factor) possible for a specific resolution, conductivity and integrator, whilst generating stable simulations and indistinguishable results for a given problem.

	\subsection{METHOD}
		All results have been generated using the METHOD code. Details of the numerical schemes used in METHOD can be found in more detail in \citep{Wright2019}. Briefly, the numerical flux is computed using flux splitting, with a third order WENO reconstruction scheme \citep{Shu1997}. Time integration uses one of two schemes: an operator-split, second-order Runge-Kutta (RK) method \citep{Gottlieb1998}; and the second-order implicit-explicit (IMEX) RK scheme from \citet{Pareschi2004}. In order to maintain the divergence constraints imposed by Maxwell's equations we use the method of hyperbolic divergence cleaning \citep{Dedner2002}.
		
	\subsection{Convergence with $\sigma$}\label{sec:convergenceWithSigma}
		We now present results to demonstrate how the new model behaves with varying magnitudes of the conductivity. A number of well-known initial data are evolved, and the results of resistive MHD and REGIME are compared and contrasted. For clarity, we have also evolved some simulations using ideal MHD to highlight its inability in modelling various resistive behaviours, and to show how the REGIME extension to it can capture these behaviours. 
		
		\subsubsection*{Brio-Wu shock tube}\label{sec:BrioWu}
			The Brio-Wu shock tube test \citep{Brio1988} is a standard numerical fluid problem used to assess how a model behaves when there is discontinuous data. The 1D domain is separated into two regions by a partition which is removed at time $T=0$. The initial data for this problem is given by $(\rho, p, B_y) = (1, 1, 0.5)$ for $x<L/2$, and $(0.125, 0.1, -0.5)$ for $x\ge L/2$, all other variables are set to zero. To make comparisons with \citep{Palenzuela2009, Dionysopoulou2013} we have set $\Gamma = 2$. The system is run until $T=0.4$, using $N_x=128$ grid points.
	
			The first three columns of figure \ref{fig:BrioWu final and error} show the final state of the $y$-direction magnetic field and pressure for ideal MHD (blue dash-dotted), resistive MHD (black dashed) and REGIME (red solid), for varying conductivities. Clearly for the larger conductivities, there is little difference in the output of the models, which is to be expected as both resistive MHD and REGIME limit to ideal MHD. For more resistive simulations (smaller $\sigma$), the resistive models differ more greatly from the ideal MHD results, and REGIME and resistive MHD results are in excellent agreement.
			
			In the final column of figure \ref{fig:BrioWu final and error}, we calculate the error as the $L_2$ norm of the difference between REGIME's output and resistive MHD, and see how this norm varies with changes in $\sigma$. Clearly, there is a significantly reduced error growth when using REGIME over the ideal MHD model---the diffusion term captures the important features coming from the resistive model.

			\begin{figure*}
				\centering
				\includegraphics[width=\linewidth]{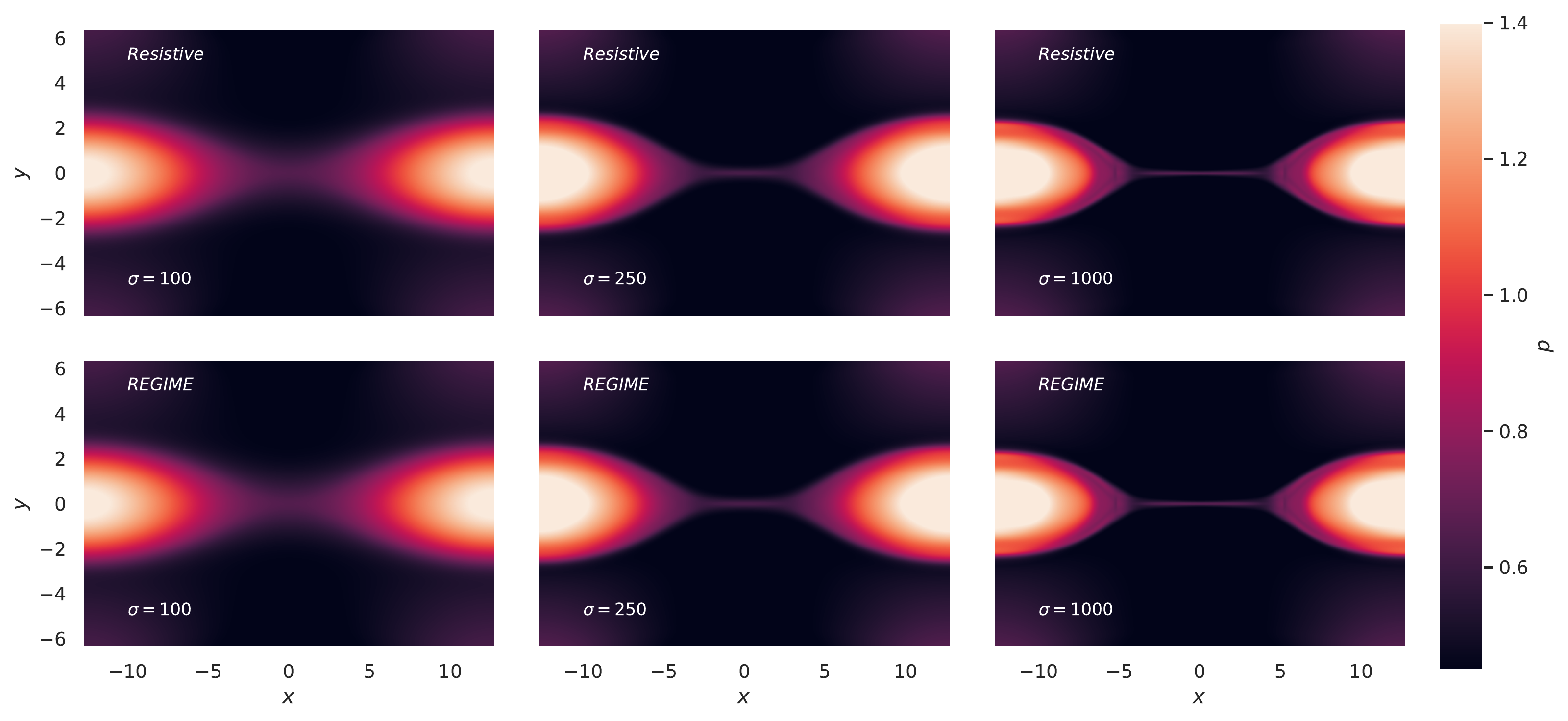}
				\caption{Final state, at $T=100$, of the pressure for the magnetic reconnection problem at a range of conductivities. The bright spots correspond to the location of the magnetic islands. We can see clearly that the more resistive systems exhibit significantly more diffusion. Once again, resistive MHD and REGIME show indistinguishable results.}
				\label{fig:RMRFinalStatePressure}
			\end{figure*}
			
		\subsubsection*{Self-similar current sheet}	
		
			As the form of the proposed source term looks much like a diffusion term, we now present a problem in which the solution obeys the diffusion equation. The self-similar current sheet test \citep{Komissarov2008} begins with a static fluid in equilibrium, and imposes upon it a magnetic field of the form $(B_x, B_y, B_z) = (0, B(x, t), 0)$. This is one of the few MHD tests in which an analytic solution exists, namely,
			\begin{align}
				B_y = B(x, t) = B_0 \ \text{erf}\bigg( \frac{\sqrt{\sigma x^2}}{2t} \bigg),
			\end{align}
			where erf is the error function and the magnetic field satisfies
			\begin{align}
				\partial_t B - \frac{\partial_x^2 B}{\sigma} = 0.
			\end{align}
			
			The system is set up with $(p, \bm{v}, \rho) = (50.0, 0.0, 1.0)$ at $T=1$, using $N_x=128$ cells and an adiabatic index of $\Gamma=2$. All simulations are run until the end time, $T=8.0$, with a magnetic field strength of $B_0 = 1$.
			
			An ideal simulation of the current sheet problem should not evolve from the initial condition, which given an infinite resolution would be a step function. Simulations with greater resistivities should expect more diffusion, and so the solution should appear more smeared out. Clearly, for the two final states (left and middle plots) from figure \ref{fig:CurrentSheet final and error}, both the resistive MHD and REGIME capture this process, with excellent agreement between both down to conductivities of only $\sigma = 20$. The error plot, rightmost in figure \ref{fig:CurrentSheet final and error}, shows little error growth between resistive MHD and REGIME.
			
		\subsubsection*{Resistive magnetic reconnection}

			Next, to demonstrate the effectiveness of REGIME in higher dimensions, we present the 2D magnetic reconnection problem. Similarly to the current sheet problem, magnetic reconnection is a purely resistive phenomenon---the lack of electrical resistivity in ideal MHD should mean that there is no evolution from the initial condition (excluding effects due to numerical diffusion). Furthermore, this kind of phenomenon should occur extensively during simulations of neutron star mergers, due to the complex interaction between the progenitors' intense magnetic fields, and so makes a useful test of REGIME in its applicability to these simulations. 
		
			Reconnection occurs when two magnetic fields whose vectors are not parallel intersect, forming a new magnetic vector and releasing stored magnetic energy into thermal and kinetic energy. There are multiple configurations of magnetic fields that will lead to some form of reconnection---indeed, understanding magnetic reconnection is so vital for solar physics that there is extensive literature and multiple models to choose from. As we are only interested in demonstrating the effect, we will adopt the first and simplest attempt of understanding reconnection, the Sweet-Parker model \citep{Parker1957, Sweet1958}. One should note, however, that \edit{the reconnection rate predicted by the Sweet-Parker model} is slow compared to other models, and experimental data. In fact, \citet{Loureiro2016} claim that ``unsatisfactory predictions are obtained for reconnection events in almost all plasmas that one cares to examine."
		
			In the Sweet-Parker model, the reconnection layer is defined to have a thickness of $2\lambda$, and a length of $2L$. The initial magnetic field strength is given by $B_0$, with the corresponding Alfven velocity as $V_A$. With this, it is possible to estimate the rate of magnetic reconnection, $\mathcal{E} \equiv u_{\text{in}}/u_{\text{out}}$. By conservation-of-mass arguments, we have the relation
			\begin{align}
				\frac{u_{\text{in}}}{u_{\text{out}}} = \frac{\lambda}{L},
			\end{align}
			where $u_{\text{in/out}}$ is the velocity of the fluid flowing in/out of the boundary layer. Using Ohm's law, one can show that the rate of flow in to the boundary layer can be expressed as $u_{\text{in}} = \eta / \lambda$, in which the resistivity is $\eta = \sigma^{-1}$. The Lundquist number, $S$, is defined via $S\equiv L V_A / \eta$ and so $\eta / V_A = L/S$. Using the fact that $u_\text{out} = V_A$, one can show that
			\begin{align}
				\mathcal{E} \equiv \frac{u_{\text{in}}}{u_{\text{out}}} = \frac{\lambda}{L} \propto \eta^{1/2}.
			\end{align} 
			
			The domain, where $x \in [-12.8, 12.8]$ and $y \in [-6.4, 6.4]$, is set up as follows:
			\begin{align}
				p &= 0.5 \\
				\rho &= \rho_{\infty} + \rho_0 \cosh^2{(y/\lambda)} \\
				B_x &= B_0 \tanh{(y/\lambda)},
			\end{align}
			with all other variables set to zero. The magnetic fields are then perturbed by, 
			\begin{align}
				\delta B_x &= - \frac{\pi \psi_0}{L_y} \sin{(\pi y / L_y)} \cos{(2 \pi x / L_x)} \\
				\delta B_y &= \frac{2 \pi \psi_0}{L_x} \sin{(2\pi x / L_x)} \cos{(\pi y / L_y)}.
			\end{align}
			Here, we have $(L_x, L_y) = (25.6, 12.8)$, with the initial thickness of the boundary layer $\lambda = 0.5$. The background density is $\rho_{\infty} = 0.2$ with $\rho_0 = 1.0$, a magnetic field strength of $B_0 = 1.0$, and a perturbation size of $\psi_0=0.1$. The simulation is then run using $1024\times512$ cells until $T=100$, with the adiabatic index as $\Gamma = 2$. We use periodic boundaries along the $x$-axis and outflow boundaries along the $y$-axis.

			In figure \ref{fig:RMRFinalStatePressure}, we show the final state of the hydrodynamic pressure for a range of conductivities for both resistive MHD and REGIME. The hotspots in the pressure coincide with the formation of two magnetic islands. Clearly, for the simulations with greater conductivity, there is significantly less diffusion occurring---this is seen as a more distinct separation of the two magnetic islands. As the resistivity of the simulations is increased, this separation reduces more quickly. With these plots, one can see no differences in the output of resistive MHD and REGIME. The proposed source term of REGIME captures the diffusive behaviour of the full model exceptionally well.
		         
			Next, to determine the rate of diffusion for a given resistivity for both models, we follow \citet{Mignone2012} and compute the current density from Ampere's law, $\bm{J} = \bm{\nabla} \times \bm{B}$, along the $y$-axis (at  $x=0$). As the rate of reconnection can be written $\mathcal{E} = \lambda / L$, where $L = L_x$ is the length of the domain in the $x$-direction, we compute the width of the boundary layer, $\lambda$, as the width of the Gaussian profile that best approximates $\bm{J}$ at $T=100.0$. 
		
		\begin{figure}
			\centering
			\includegraphics[width=\linewidth]{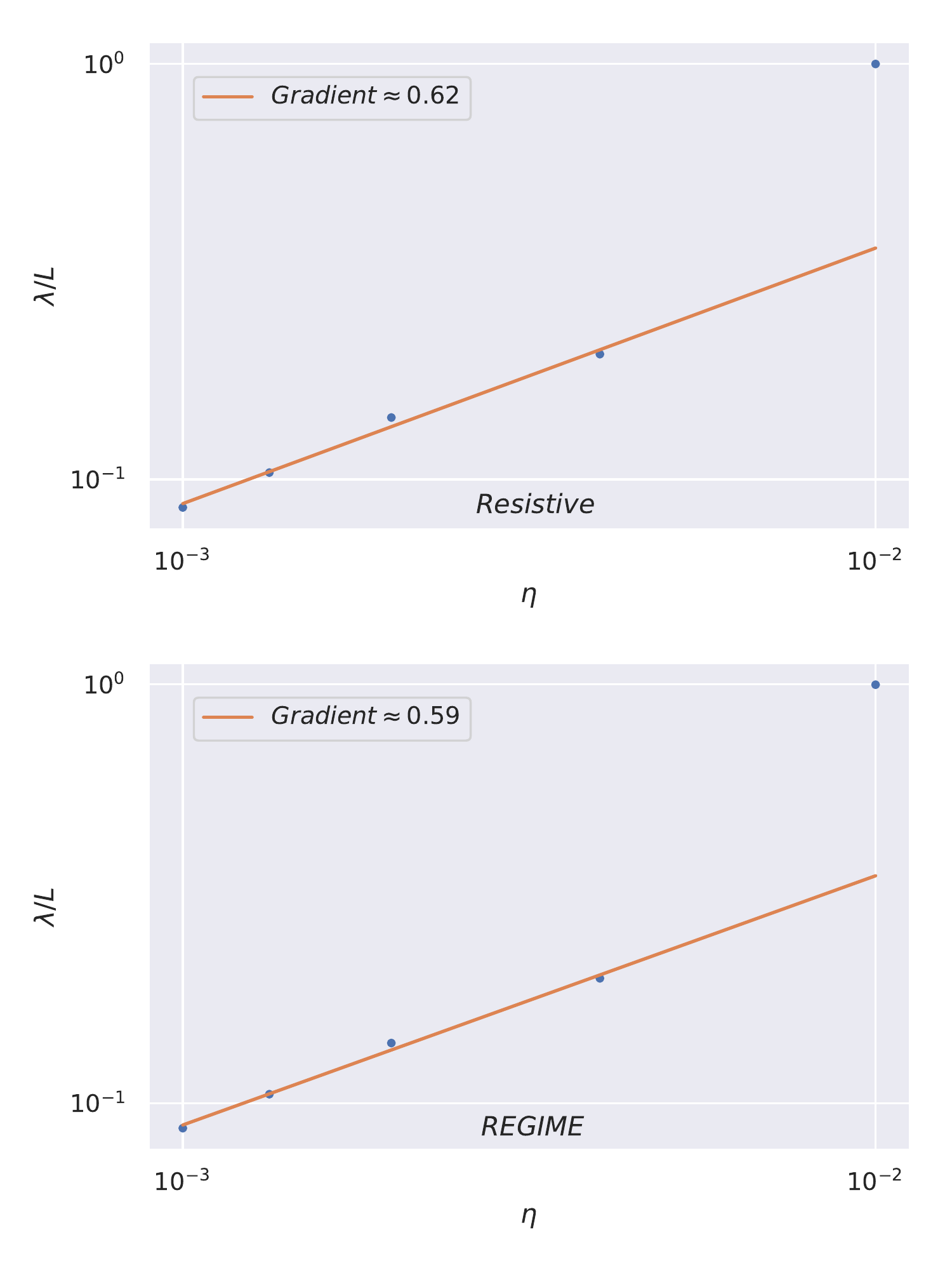}
			\caption{Using the width of the best fit Gaussian, we plot here the reconnection rate for a range of resistivities for the resistive MHD model (top) and REGIME (bottom). The gradient is highlighted in the legend, and shows the models' reconnection rate agree well with each other and the Sweet-Parker prediction.}
			\label{fig:RMRScalingLawFinal}
		\end{figure}
			As the fit for $\sigma=100$ is poor, it has been excluded from the reconnection rate analysis. The results of this analysis are shown in figure \ref{fig:RMRScalingLawFinal} for a number of resistivities, and shows a scaling slightly greater than what is expected from the Sweet-Parker analysis for both resistive MHD and REGIME. Clearly the subgrid source term captures this effect, and as a result the reconnection rate of the two models agree remarkably well.

		\subsubsection*{Kelvin-Helmholtz instability}\label{sec:KHI}
			
			A common type of fluid instability that is believed to play an important role in evolution of magnetic fields during mergers \citep{Kiuchi2015} and other astrophysical events is the Kelvin-Helmholtz instability (KHI). In this scenario, two differentially flowing fluids are perturbed along the boundary layer between them, resulting in the growth of vortices and a corresponding cascade of kinetic energy from large scales down to the smallest dynamos. The motion of the fluid couples to the magnetic fields, where the folding and twisting increases the fields' energy density and strength. This is one mechanism by which magnetars are thought to develop such intense magnetic fields.

			\begin{figure}
				\centering
				\includegraphics[width=\linewidth]{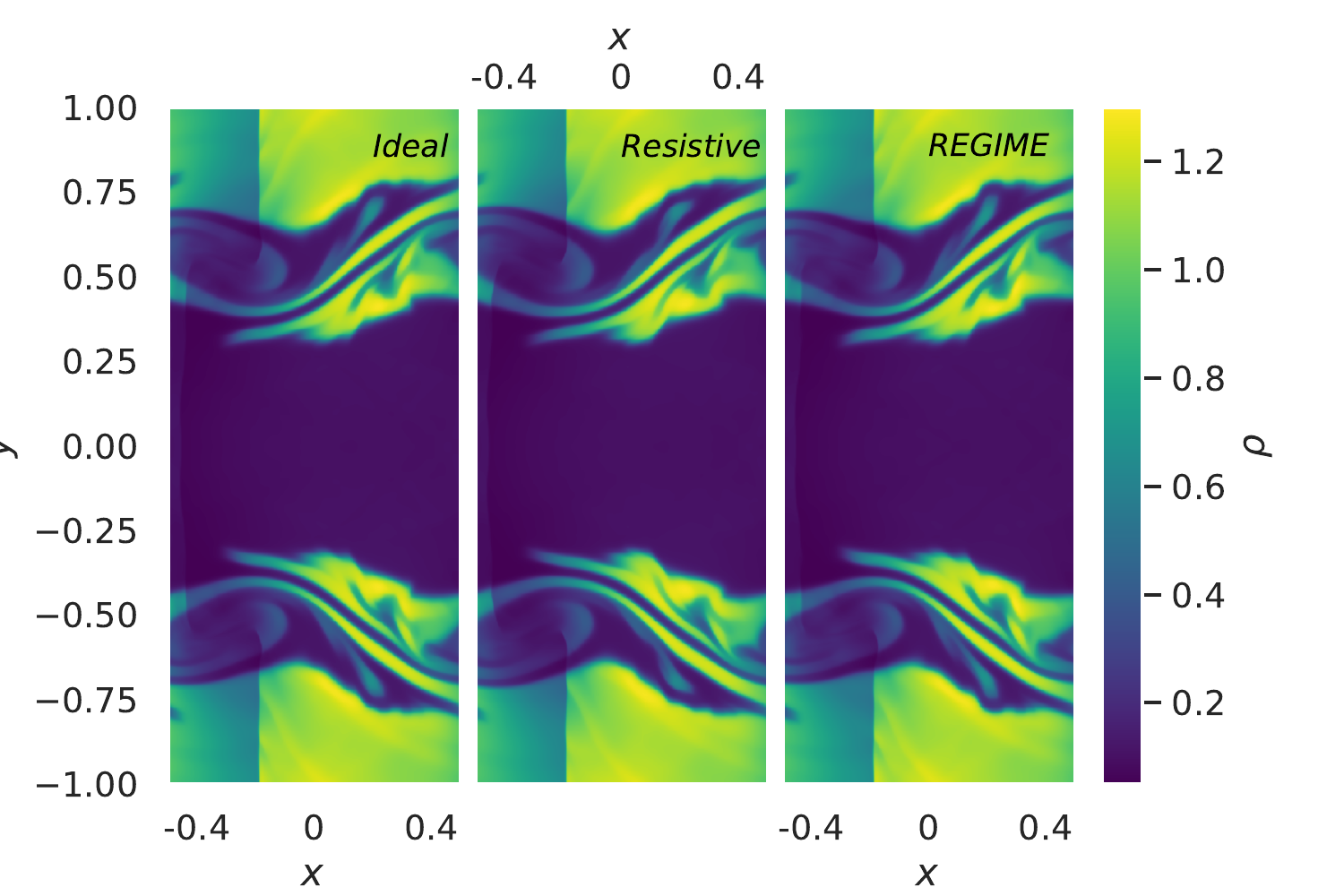}
				\caption{Final state, $T=6.0$, of the density for the Kelvin-Helmholtz instability for each model. Visually, there is no difference between the solutions.}
				\label{fig:KHIDensityFinal}
			\end{figure}
			
			To investigate this process, we use the initial conditions from \citep{Beckwith2011}. We simulate a 2D domain where $x \in [-0.5, 0.5]$ and $y\in[-1.0, 1.0]$, with the initial $x$-velocity as

			\begin{align*}
				v_x &= 
				\begin{cases}  \label{eq: KHI vy single fluid}
					&\ \ v_{\text{shear}} \tanh \big(\frac{y-0.5}{a} \big)	\ \ \ \ \ \ \ \ \ \ \ \ \ \ \text{if} \ \ y > 0.0 \\
					&-v_{\text{shear}} \tanh \big( \frac{y+0.5}{a} \big) \ \ \ \ \ \ \ \ \ \ \ \ \ \ \text{if} \ \ y \le 0.0, \\
				\end{cases} 
			\end{align*}
			the density as
			\begin{align*}
				\rho &= 
				\begin{cases}
					&\rho_0 + \rho_1 \tanh \big(\frac{y-0.5}{a} \big) \ \ \ \ \ \ \ \ \ \ \ \ \ \ \  \text{if} \ \ y > 0.0 \\
					&\rho_0 - \rho_1 \tanh \big( \frac{y+0.5}{a} \big) \ \ \ \ \ \ \ \ \ \ \ \ \ \ \ \text{if} \ \ y \le 0.0, 
				\end{cases}
			\end{align*}
			and then perturb the $y$-velocity as
			\begin{align*}
				v_y &= 
				\begin{cases}
					&\ \ A_0 v_{\text{shear}} \sin (2 \pi x) \exp^{\frac{-(y-0.5)^2}{l^2}}	\ \ \ \ \text{if} \ \ y > 0.0 \\
					&-A_0 v_{\text{shear}} \sin (2 \pi x) \exp^{\frac{-(y+0.5)^2}{l^2}}	\ \ \ \ \text{if} \ \ y \le 0.0, \\
				\end{cases} 
			\end{align*}
			where the shear velocity is $v_{shear}=0.5$, a boundary layer thickness of $a=0.01$, the densities are given by $(\rho_0, \rho_1) = (0.55, 0.45)$, and a perturbation amplitude of $A_0=0.1$ over a length $l=0.1$. The initial pressure is uniform, $p=1.0$, and we impose a perpendicular (to the flow) magnetic field of $B_z=0.1$. The adiabatic index is set to $\Gamma = 4/3$ and the simulation is evolved until after the end of the linear growth stage, at $T=6.0$, using $(N_x, N_y) = (512, 1024)$. As in resistive reconnection, we use periodic boundaries in the $x$-direction and outflow boundaries in the $y$-direction.

\begin{figure}
				\centering
				\includegraphics[width=\linewidth]{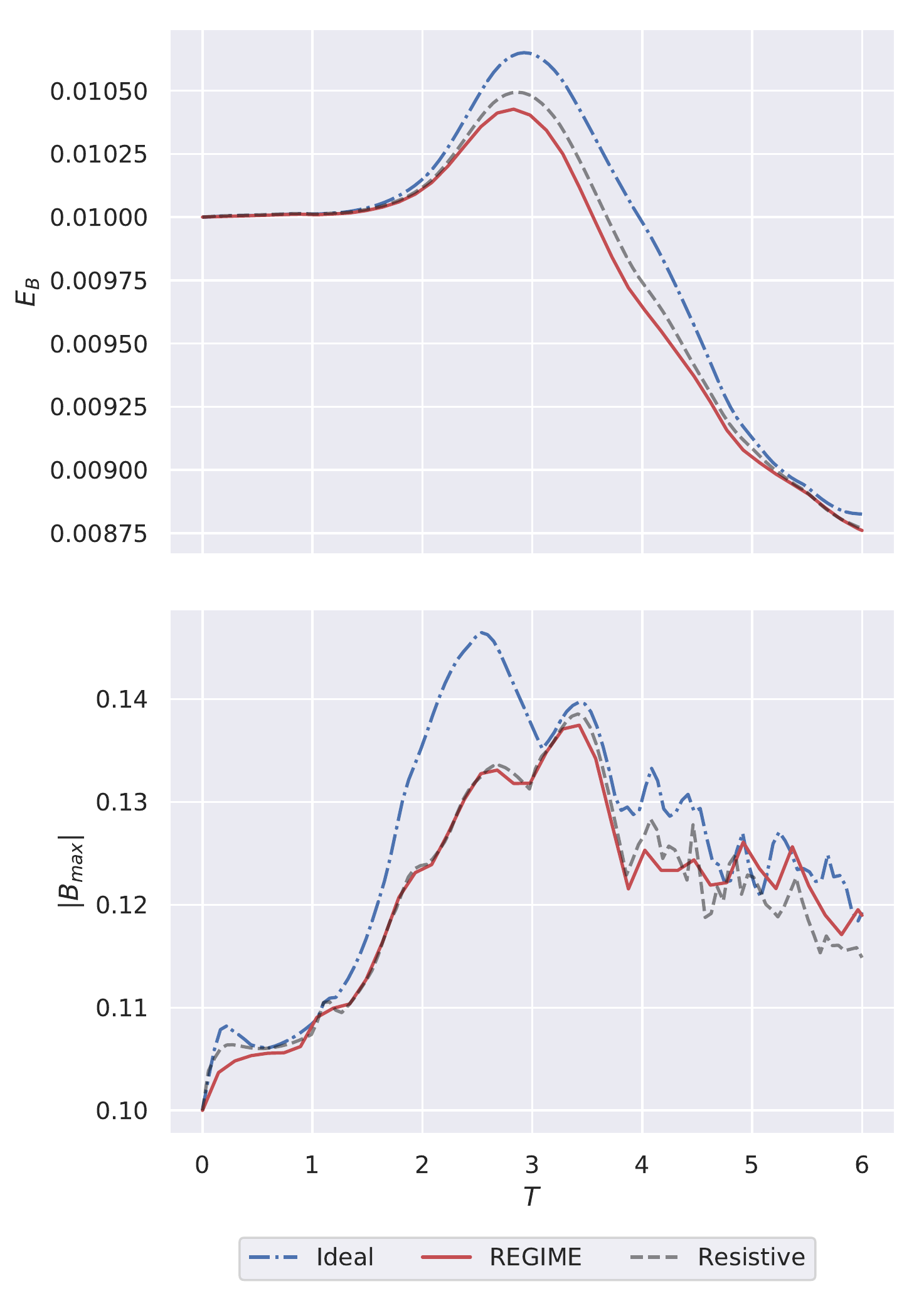}
				\caption{Large-scale magnetic evolution for the Kelvin-Helmholtz instability. Top: the average magnetic energy density across the whole domain. Bottom: the maximum magnetic field strength. From these images, we can see that REGIME modifies the behaviour of the magnetic fields from the ideal case, replicating the behaviour of the resistive model very well.}
				\label{fig:magnetic evolution}
			\end{figure}

			Whilst there is little-to-no difference between the observed state in the density between ideal MHD, resistive MHD or REGIME, figure \ref{fig:KHIDensityFinal}, we can be more quantitative. As the evolution of the magnetic fields can be affected by the dynamics of this instability, in figure \ref{fig:magnetic evolution} we show their large scale behaviour. In the first plot, we show the average magnetic energy density across the whole domain, and the maximum magnetic field strength in the second.

			In ideal MHD, the magnetic field lines lock to the fluid motion, a phenomenon known as magnetic freezing, and as such are efficiently folded within features such as vortices. In resistive MHD however, the magnetic fields are free to move, leading to less efficient folding and a reduced amplification. This effect is clearly captured by REGIME on account of the reduction in the average and maximum magnetic field strength.
			
			We can also determine how REGIME behaves across all scales in the simulation. Kolmogorov predicted \citep{Qian1994} what the distribution of energy should look like across all scales for fully developed turbulence, known as his $5/3$-law. Through dimensional arguments alone, he demonstrated that the power spectrum of the kinetic energy density should satisfy
			\begin{align}
			P_{T}(k) \propto k^{-5/3},
			\end{align}
			where $k$ is the wave number of the mode.
			
			\begin{figure}
				\centering
				\includegraphics[width=\linewidth]{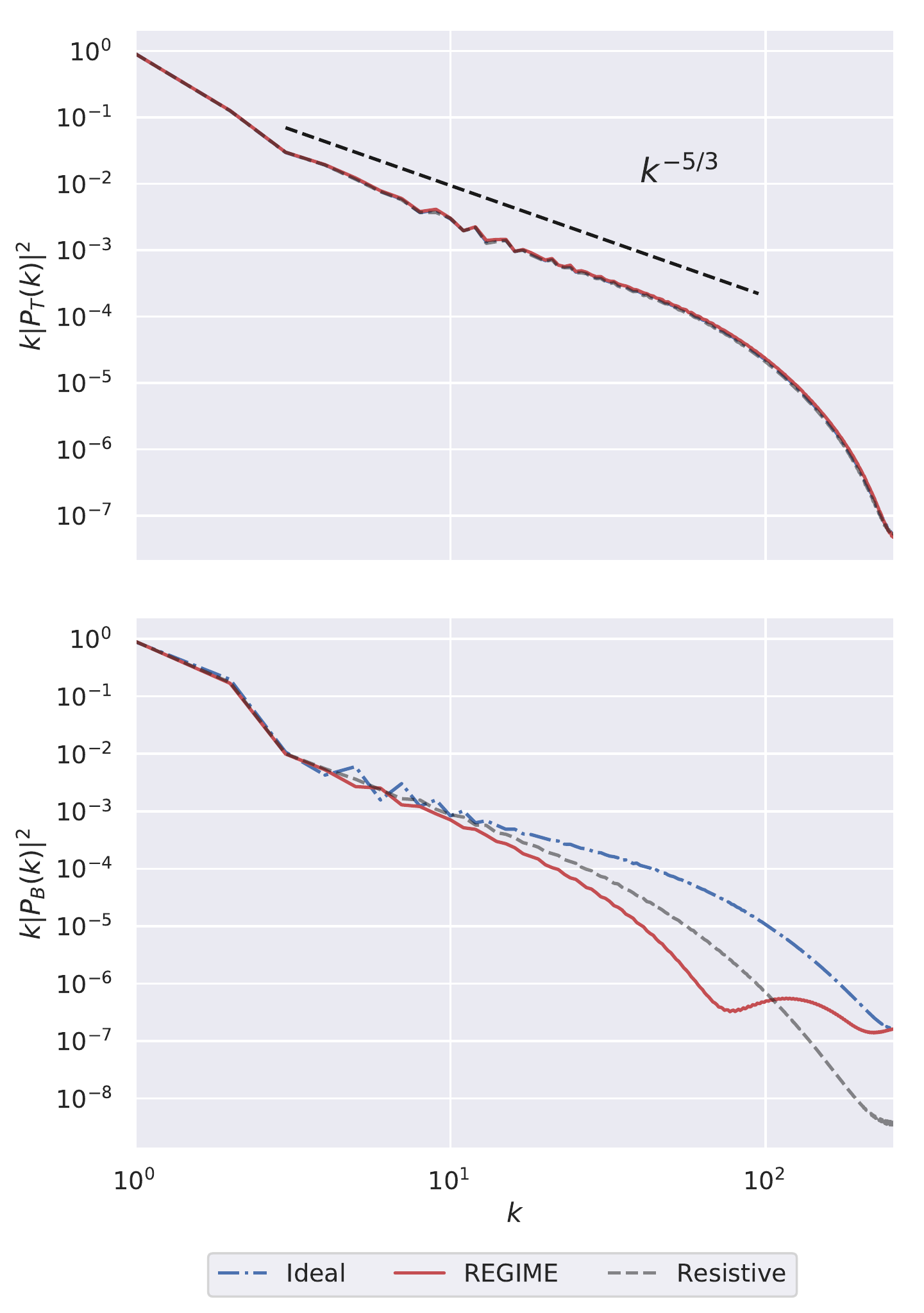}
				\caption{Power spectrum for the kinetic energy density (top) and magnetic energy density (bottom) at $T=3$. For larger wavenumbers, the magnetic energy density power spectrum of REGIME begins to deviate from resistive MHD.}
				\label{fig:KEPowerSpec}
			\end{figure}
			Using the procedure laid out in \citet{Beckwith2011}\footnote{The scripts used for this analysis can also be found at the METHOD GitHub page.}, we compute the kinetic energy density power spectrum, and also the spectrum for the magnetic energy density, comparing the results for ideal and resistive MHD and REGIME. The results are given in figure \ref{fig:KEPowerSpec}. We can see that for the kinetic energy density spectrum, all models exhibit the predicted behaviour.
			
			For the magnetic energy density power spectrum, however, we can see differences in how the models behave on small scales (large $k$). There is less energy transferred to the smaller scales for resistive MHD than for ideal MHD. This is due to no magnetic freezing, as discussed before.
			
			REGIME seems to be able to capture this change in the dynamics down to a wavenumber of $k \approx 40$, but for larger modes (smaller scales) transfers too much energy to the magnetic fields. The fraction of energy that this represents in this simulation is minute, and as a result the dynamics of the simulation remain relatively unchanged.

	\subsection{Stability}\label{sec:stability results}
		We now turn our attention to the stability of the new model. In Section \ref{sec:stability theory}, we argue that there should be some relationship between the resolution and the system's stability, and that the result should be some maximum possible spatial resolution whilst still giving stable evolutions.

	\begin{figure}
		\centering
		\includegraphics[width=\linewidth]{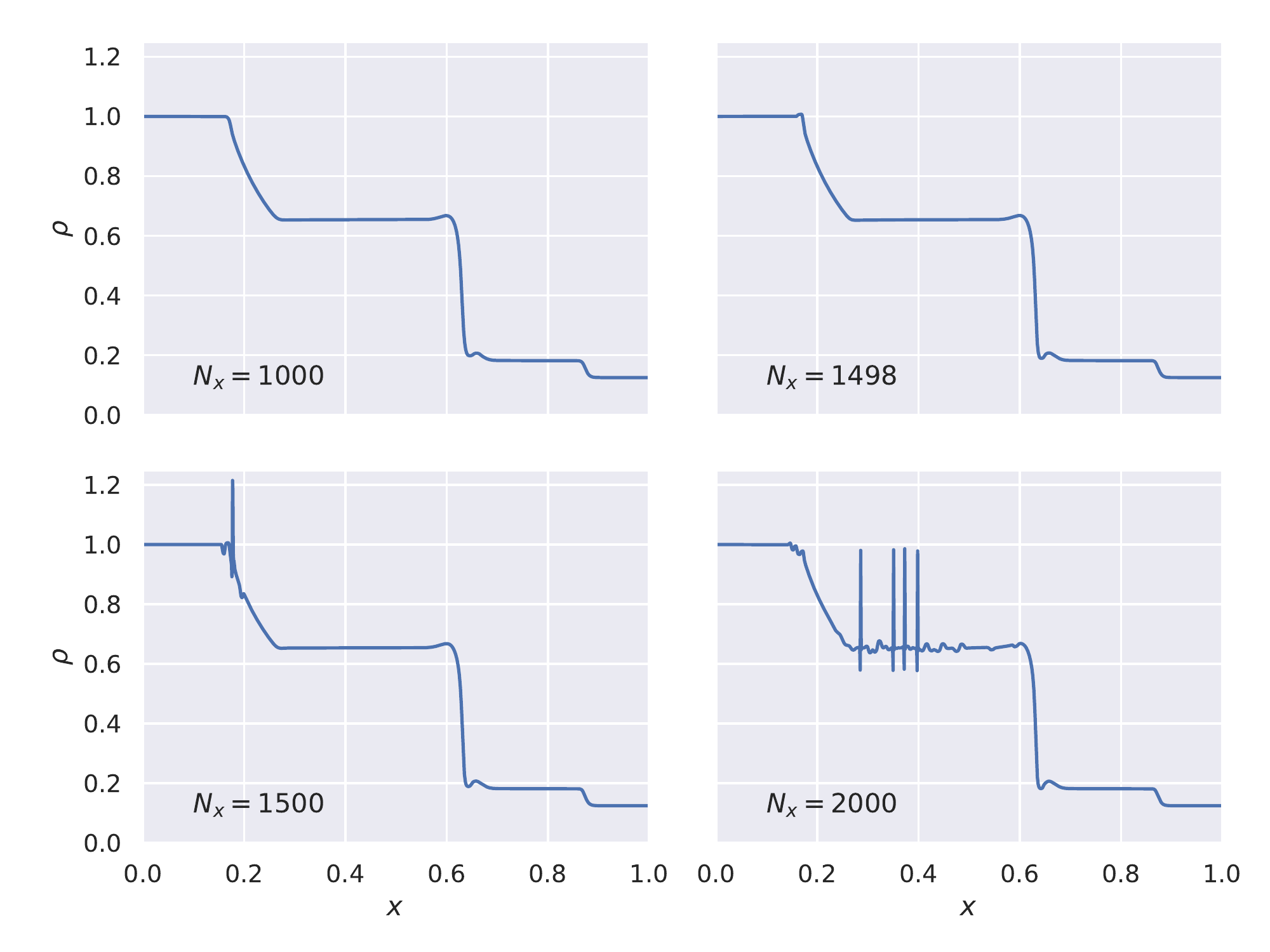}
		\caption{Final state of the density for the Brio-Wu shock-tube problem with increasing resolution. For simulations with $N_x>1498$ the stability criterion is violated and the evolution becomes unstable.}
		\label{fig:BrioWuFinalStability}
	\end{figure} 

		Returning to the Brio-Wu shock tube test, section \ref{sec:BrioWu}, we can see how quickly the simulation can move from a stable region to an unstable region. Figure \ref{fig:BrioWuFinalStability} shows the final state of the density for the Brio-Wu shock tube, where $\sigma=1000$, for four different resolutions. The first two (top row) use $N_x = 1000$ and $1498$ cells, leading to stable simulations. Adding only two additional cells, however, results in an unstable simulation, as can be seen by the spikes present in the data at $x \approx 0.2$. If we continue to use finer resolutions, the effect becomes more pronounced, and additional artefacts appear.

		\begin{figure*}
			\centering
			\includegraphics[width=\linewidth]{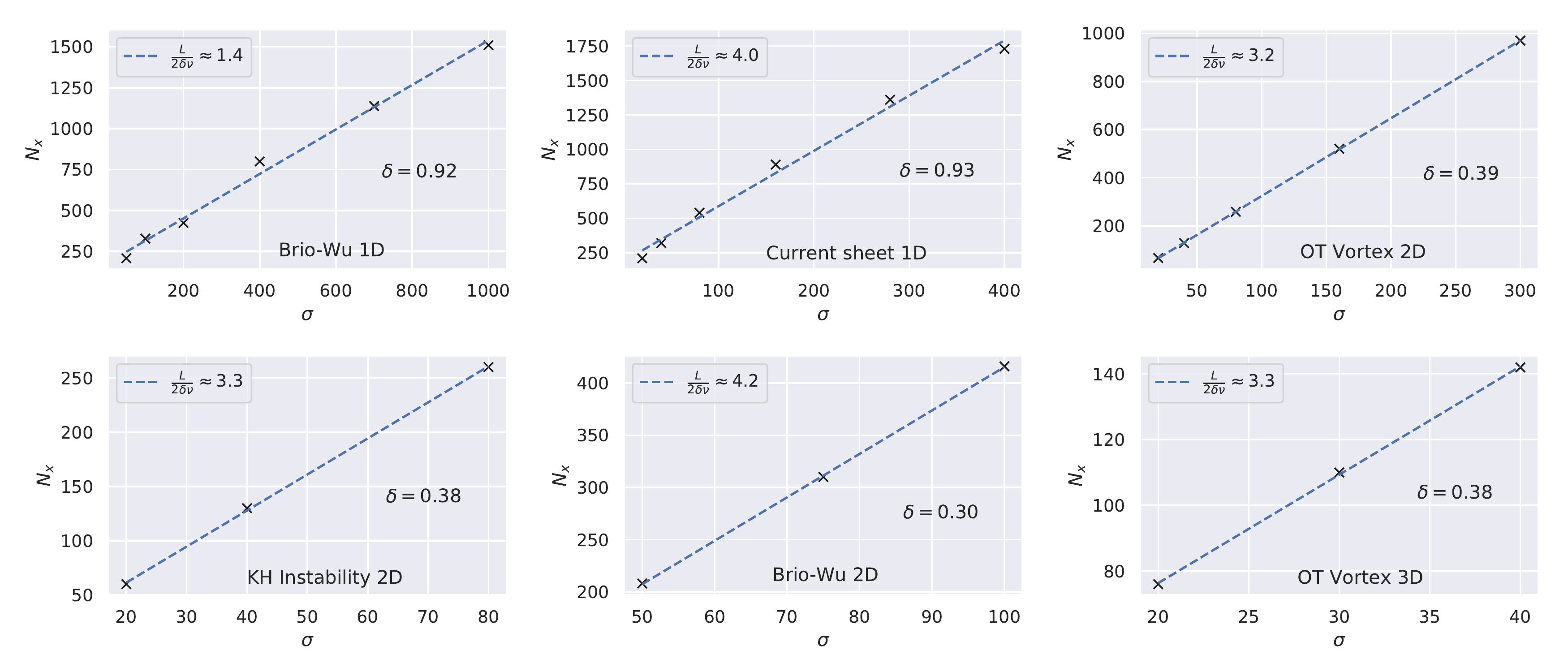}
			\caption{Stability criterion for 1D, 2D and 3D test problems. Points represent the resolution of the last stable simulation for the indicated initial data, with the gradient highlighted in the legend, and the corresponding value for the softening parameter, $\delta$. This suggests that the criterion, equation \ref{eq:stability criterion}, should unsure stability when satisfied with a softening factor of $\delta = 1$.}
			\label{fig:BWandCSDelta}
		\end{figure*}
	
		When performing the stability analysis, we assumed that in moving to higher dimensions it was the same largest eigenvalue that is the main factor in determining stability. To test this, we perform a number of simulations in all dimensions. The initial data for the Orszag-Tang vortex is taken from \citet{Beckwith2011} for 2D, with the addition of a $z$-perturbation to the fluid velocity of the form $0.2 \sin(2 \pi z)$ for the three dimensional case.
		
		Figure \ref{fig:BWandCSDelta} shows the relationship between the finest possible resolution for a range of conductivities. The one-dimensional Brio-Wu and current sheet problems produce an average $\delta \approx 0.92$ and $0.93$, respectively. When we move to higher dimensions, the $\delta$ value for all problems drops to a value of $\delta \approx 0.3-0.4$, corresponding to a less restrictive stability criterion. 
		
		Whilst the reasoning behind this drop in $\delta$ is not fully understood, we believe it comes from the numerics. The fifth plot (bottom left) shows the data for a 2D Brio-Wu test, in which the standard 1D data is rotated about the $z$-axis by $\pi/4$. If the stability was set by the 1D nature of the initial data, a rotation about the $z$-axis should not change the resolution that is possible along the propagating shock wave, and we would expect the same softening factor. In reality, the rotated problem allows for higher resolutions to be achieved, possibly due to the effects of the cross derivatives in the REGIME source term.

		Whatever the reason, the effect is to allow a greater resolution in the multi-dimensional case, tending towards a factor $3\times$ for large conductivities. If we therefore take a conservative value for the softening factor of $\delta = 1$, we can determine whether the types of astrophysical simulation we are interested in are possible with REGIME. The conductivities one would expect in a warm neutron star crust have values of $\sigma \sim 10^{20-22}$\,s\,$^{-1}$ \citep{Harutyunyan2016}. Such a conductivity would allow resolutions on the order of millimetres\footnote{We have $\Delta x > 2 \delta \nu / \sigma$. When redimensionalising we require a factor of $c^2$ on the RHS, thus $\Delta x > 2 \delta \nu c^2 / \sigma$.}, whereas the highest resolution simulations to date \citep{Kiuchi2014} have resolutions of tens of metres in the central region. Furthermore, the conductivities used in \citet{Carrasco2019} to model the exterior of a magnetar quote conductivities of $\sigma \sim 10^{17}$\,s\,$^{-1}$, resulting in a maximum resolution of metres. As a result, it appears that REGIME will be suitable for realistic neutron star simulations, and that for physical magnitudes of the conductivity it should be stable, even for the finest resolutions to date.

	\subsection{Performance}\label{sec:performance}
		Knowing that REGIME gives sensible results for a range of simulations, and agrees well with resistive MHD in various regimes, we now turn our attention to the relative performance of the two models. The motivation for the model is that resistive MHD simulations generally require implicit time-integration schemes to ensure stable evolutions, and as a result suffer from slow evolution especially when near the ideal MHD limit (large $\sigma$). By instead using the REGIME source term, which is small in this limit compared to the resistive MHD source term, it is possible to perform the evolution using explicit time integrators, and ultimately simulations should benefit in terms of performance.
		
		To determine this, we once again return to the Brio-Wu test, and to the magnetic reconnection problem. The Brio-Wu shock-tube is run using $N_x=400$ cells and is run until $T=0.4$. The 2D reconnection problem uses $(N_x, N_y) = (128, 64)$ run until $T=2.0$. Due to time limitations the 3D reconnection problem uses $(N_x, N_y, N_z) = (128, 64, 64)$, and is run until $T=0.5$. The problems are evolved using both models for a range of conductivities, and the optimum Courant factor giving the fastest execution, whilst generating similar results, is used. For the resistive model, we present the execution time using both the operator-split RK2 method and the SSP2(222) IMEX \edit{\citep{Pareschi2004}} integrator for comparison. Conductivities that have no corresponding execution time failed to run even with unreasonably small Courant factors.
		\begin{figure*}
			\centering
			\includegraphics[width=\linewidth]{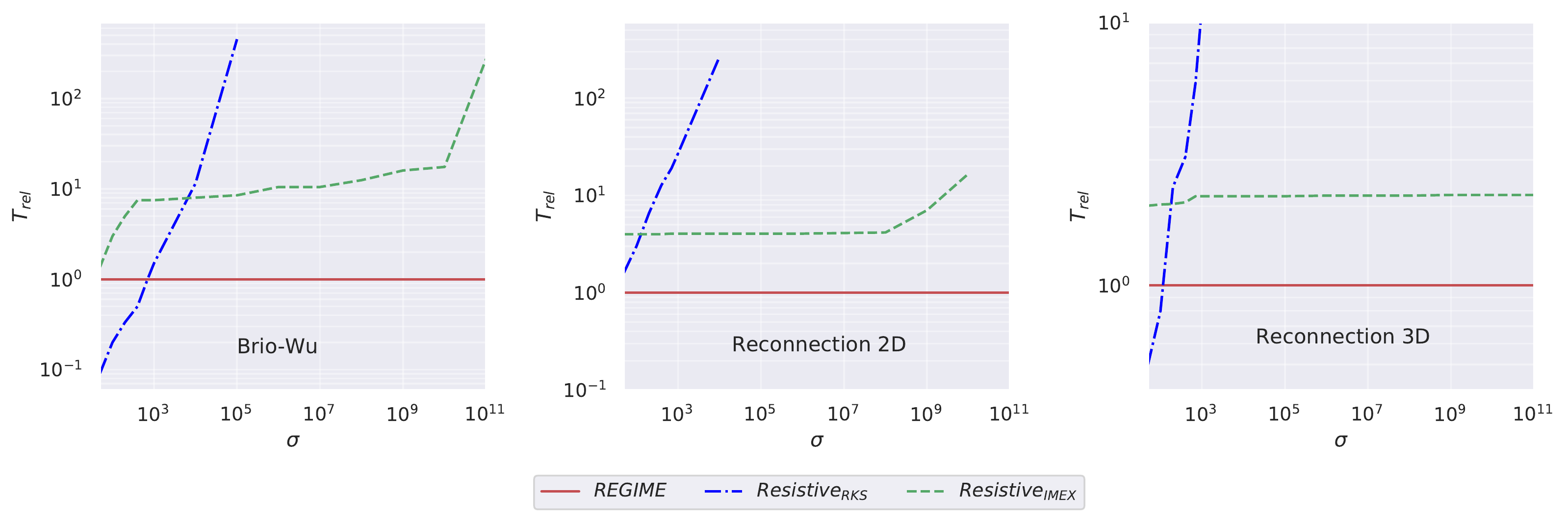}
			\caption{The relative execution time for the Brio-Wu and magnetic reconnection (2D and 3D) test problems---execution times are relative to REGIME. The red line corresponds to REGIME using the RK split integrator, and the blue and green line represent resistive MHD with the RK split and IMEX integrators, respectively. Using REGIME always results in a more efficient simulation than resistive MHD when using the IMEX integrators. \edit{Units for $\sigma$ here are in natural code units, as in the rest of the paper.}}
			\label{fig:BWandReconnectionOptPerf}
		\end{figure*}
		
		Results are given in figure \ref{fig:BWandReconnectionOptPerf} that demonstrate the performance benefits of REGIME over resistive MHD. For the majority of the range of $\sigma$ that we present here, REGIME is the most efficient model to evolve for both problems, being only second to the resistive model (using explicit integrators) for the most resistive set-ups.
		
		With these results, we can see that for any given resistivity there is no need to employ implicit integrators to evolve the system---either the system has a small enough $\sigma$ to allow use of the explicit integrators to evolve resistive MHD, or REGIME will provide a faster execution when implicit schemes are required.

\section{Discussion}\label{sec:discussion}

	We present a resistive extension to the relativistic, ideal MHD equations often used in astrophysical simulations. The new source term is derived by expanding equations of motion about an equilibrium solution, resulting in a perturbation term that exhibits diffusive behaviour. The diffusion term scales with the resistivity, $\eta \equiv \sigma^{-1}$, and is numerically non-stiff when the resistive MHD equations are stiff. As a result, near the ideal limit, the source term may be evolved explicitly, resulting in speed-ups in execution times of many factors.
	
	The new source term is demonstrated to produce similar results to resistive MHD in a range of initial set-ups, is able to capture resistive effects near discontinuous data without the onset of Gibbs oscillations, and shows little error growth for smooth, resistive solutions over conductivities spanning many orders of magnitude. For more complex simulations of magnetic reconnection, the correct scaling laws are produced for the reconnection rate. 
	
	Because of the nature of the source term, the explicit evolution of the new system results in relatively quick executions. Runtimes can vary dependent upon the conductivity but range from $2$ to $10^2\times$ faster than the optimum evolution using either explicit or semi-implicit integrators with resistive MHD. As a result, for many types of simulations, there is no need to employ IMEX integrators to ensure stability with resistive MHD. Near-identical results can be produced in a fraction of the time using the REGIME source term, combined with standard, operator split RK methods.
	
	A limitation on the maximum resolution possible for a given simulation is given, and we show how the new source term can produce instabilities when this criterion is not met. We also give the form to predict at which resolutions such instabilities may occur for a given simulation, and conclude that REGIME should be stable for realistic resistivities in neutron stars.
	
	The only significant difference between REGIME and resistive MHD is seen in the magnetic energy density power spectrum of the Kelvin-Helmholtz instability. Whilst the model still follows Kolmogorov's 5/3 law for the kinetic energy power spectrum, an analysis of the magnetic energy density power spectrum shows that there are discrepancies between the expected result of resistive MHD and REGIME on small scales. These differences are not enough to affect the large scale behaviour of the magnetic fields, however, and the correct evolution of the maximum and average magnetic field is recovered.

    \edit{One outstanding challenge for simulations of NS-NS and NS-BH mergers is to accurately model both the highly conducting region within the neutron star, and consistently match the electromagnetic fields to the near-vacuum exterior. Within the neutron star, ideal MHD is expected to be a good approximation. In the exterior, there are two possible limits of the resistive MHD equations: the force-free approximation ($\rho \to 0, \sigma \to \infty$) or the electro-vacuum ($\rho, \sigma \to 0$). In both cases we need to consider the conductivity $\sigma$ varying with space and time. In principle the REGIME approach extends directly to the varying-$\sigma$ case. However, by construction, it will become stiff in the electro-vacuum ($\sigma \to 0$) limit, and hence not practical. Current approaches, as proposed in \citep{Lehner2012, Dionysopoulou2013}, expect that in the regions around the merger there would be sufficient matter such that the force-free approximation would be valid. In this limit, therefore, REGIME should be sufficient in capturing the highly conducting plasma. If the $\sigma \to 0$ limit is more suitable then it would be more practical to match to a fully resistive model in the exterior, as it would not be numerically stiff in that case, whilst evolving the interior using REGIME. This would be analogous to the approach used in \citep{Lehner2012}.}
	
	\edit{In Section \ref{sec:numerics}, to simplify the form of the inverted matrices in the source, we made the assumption that the fluid only couples weakly to the magnetic fields, and thus set $E=B=0$ for those calculations. In the tests we performed in this paper, this simplification gave only minor differences in the output of REGIME when compared to a full, numerical inversion. One would, however, expect these differences to grow if one modelled a magnetically dominated (low $\beta$) plasma.}
	
	Although all simulations have been performed in the special relativistic limit, the techniques we have used are not limited to this alone. A general relativistic extension to REGIME is under way, and will soon allow for more efficient, resistive simulations of neutron star mergers, accretion on to compact objects, and magneto-rotational instabilities. 
	
	\section*{}
	
	The authors acknowledge the use of the IRIDIS High Performance Computing Facility, and associated support services at the University of Southampton, in the completion of this work. We also gratefully acknowledge financial support from the EPSRC Centre for Doctoral Training in Next Generation Computational Modelling grant EP/ L015382/1. Open source software used includes Matplotlib \citep{Hunter2007}, Seaborn \citep{Waskom2014} and CMINPACK \citep{Devernay2017}.






\bsp	
\label{lastpage}
\end{document}